\documentclass[useAMS,usenatbib]{aa}

\usepackage{xcolor}
\usepackage{graphicx}
\usepackage{subcaption}
\usepackage{orcidlink}

\newcommand{\fr}[1]{\begin{color}{red}\ \ifthenelse{\equal{#1}{}} {$\bullet\bullet\bullet$} {$\bullet$\ #1 $\bullet$}\end{color}}

\begin{document}

\title{Asymmetries in stellar streams induced by a galactic merger}

\author{Claire~Guillaume\thanks{\email{claire.guillaume@astro.unistra.fr}}\inst{\ref{sxb}}\orcidlink{0009-0005-6540-0229}
\and Florent~Renaud\inst{\ref{sxb},\ref{usias}}\orcidlink{0000-0001-5073-2267}
\and Nicolas~F.~Martin\inst{\ref{sxb},\ref{MPI}}\orcidlink{0000-0002-1349-202X}
\and Benoit~Famaey\inst{\ref{sxb}}\orcidlink{0000-0003-3180-9825}
\and Paola~Di~Matteo\inst{\ref{LIRA}}\orcidlink{0000-0002-5213-4807}
\and Guillaume~F.~Thomas\inst{\ref{IAC},\ref{Uni_Laguna}}\orcidlink{0000-0002-2468-5521}
\and Salvatore~Ferrone\inst{\ref{LIRA},\ref{Dep_Phy_Rome},\ref{ICS}}\orcidlink{0000-0003-1623-6643}
\and Rodrigo~Ibata\inst{\ref{sxb}}\orcidlink{0000-0002-3292-9709}
\and Giulia~Pagnini\inst{\ref{sxb}}
}

\institute{Observatoire Astronomique de Strasbourg, Universit\'e de Strasbourg, CNRS UMR 7550, F-67000 Strasbourg, France\label{sxb} 
\and University of Strasbourg Institute for Advanced Study, 5 all\'ee du G\'en\'eral Rouvillois, F-67083 Strasbourg, France\label{usias}
\and Max-Planck-Institut für Astronomie, Königstuhl 17, 69117 Heidelberg, Germany\label{MPI}
\and Instituto de Astrof\'isica de Canarias, E-38205 La Laguna, Tenerife, Spain\label{IAC}
\and Universidad de La Laguna, Dpto. Astrofísica, E-38206 La Laguna, Tenerife, Spain\label{Uni_Laguna}
\and LIRA, Observatoire de Paris, Universit\'e PSL, Sorbonne Universit\'e, Universit\'e Paris Cit\'e, CY Cergy Paris Universit\'e, CNRS,
92190 Meudon, France\label{LIRA}
\and Department of Physics, University of Rome “La Sapienza”, Piazzale Aldo Moro 5, 00185 Rome, Italy\label{Dep_Phy_Rome}
\and Institute for Complex Systems CNR, Piazzale Aldo Moro 2, 00185 Rome, Italy\label{ICS}
}
\date{Received 3 October 2025 / Accepted 17 November 2025} 

\abstract {Stellar streams are sensitive to perturbations from, e.g., giant molecular clouds, bars and spiral arms, infalling dwarf galaxies, or globular clusters which can imprint gaps, clumps, spurs, and asymmetries in tails.  In addition to these effects, the impact of a galactic major merger on a population of stellar streams remains to be explored. Here, we focus on the emergence and longevity of asymmetries between the leading and trailing tails of streams caused by such interactions. We run collisionless N-body simulations of a Milky Way–like galaxy hosting 36 globular cluster streams and merging with a perturber galaxy. We propose a new asymmetry metric to quantify the structural differences between both tails from their respective cumulative density profiles. We find that the over- and under-densities along streams induced by the merger depend on the orbital characteristics of their progenitors. The non-simultaneity of this effect from stream to stream implies that global asymmetry signatures are less prominent than in individual cases. These population-averaged imprints remain detectable over only 2\,Gyr but asymmetric signatures can persist over much longer periods for individual streams with wide orbits that have been perturbed prior to coalescence. We thus caution that the interpretation of streams' morphology in the context of dark matter mapping is strongly subject to degeneracies and should be performed considering the merger history of the host.}


\maketitle

\section{Introduction}

Stellar tidal streams, the elongated tails of stars stripped from star clusters or dwarf satellites, are among the few coherent structures in the Galactic halo and serve as powerful probes of the Milky Way’s mass distribution \citep{Gibbons2014, Ibata2024, Bayer2025}. Their long and dynamically cold nature makes them highly sensitive to gravitational perturbations, providing a unique way to possibly detect otherwise invisible dark matter subhaloes \citep{Ibata2002, Johnston2002, Yoon2011}. Characteristic features such as gaps and spurs in the density and morphology of stellar streams provide constraints on the mass spectrum and abundance of dark matter subhaloes, offering a critical test of $\Lambda$CDM predictions on small scales \citep{Erkal2016, Bonaca2019}.

In addition to dark matter subhaloes, several baryonic structures within the Milky Way can also perturb stellar streams. The galactic bar and spiral arms can induce broad fans of tidal debris and generate wide gaps along the stream \citep{Pearson2017, Erkal2017, Banik2019} as well as stream-orbit misalignment that depend on the pattern speeds of the perturbers \citep{Thomas2023}. Giant molecular clouds (GMCs) can produce density variations that closely resemble those caused by dark matter subhaloes \citep{Amorisco2016, Erkal2017, Banik2019}, while close passages of globular clusters have recently been shown to create distinct gaps in streams \citep{Ferrone2025}. Epicyclic overdensities may likewise be mistaken for signatures of interactions with dark matter subhalos \citep{Ibata2020}. Beyond the $\Lambda$CDM framework, modified gravity scenarios such as MOND can also produce asymmetric tidal tails in streams \citep{Thomas2018, Kroupa2024, Pflamm-Altenburg2025}. These studies often adopt a static background gravitational potentials for the Milky Way, or use $N$-body realizations of the Galaxy run in isolation.  While such models have been instrumental in identifying the mechanisms responsible for stream perturbations, they omit a crucial aspect of Galactic evolution: its hierarchical assembly.

In the cosmological context, galaxies like the Milky Way are built through successive mergers with smaller structures \citep{White1978}. The Milky-Way is believed to have experienced a massive early merger, approximately 10\,Gyr ago, known as Gaia-Sausage-Enceladus \citep{Helmi2018, Belokurov2018} followed by a more quiescent period of minor accretions \citep{Naidu2020, Helmi2020, Kruijssen2020, Malhan2022}. Currently, the Milky Way is interacting with several massive satellites, including the Large and Small Magellanic Clouds (LMC and SMC) \citep{Besla2007, Penarrubia2016} and the progenitor of the Sagittarius stream \citep{Ibata1994}. While interactions with the LMC and the Sagittarius dwarf spheroidal galaxy are known to affect stellar streams in the MW halo (see \cite{Dillamore2022} for Sagittarius and \cite{Erkal2019}, \cite{Vasiliev2021}, \cite{Koposov2023} or \cite{Brooks2025} for LMC), the impact of past galactic mergers on stellar streams remains largely unknown. A key question is whether pre-existing streams at the time of a merger could have survive and, if so, whether they would retain signatures of the event. If such imprints persist, they could provide a unique opportunity to reconstruct the assembly history of our galaxy. \citet{Weerasooriya25} have begun to explore these questions by contrasting pre- and post-merger states, but a full time-domain treatment across the simulation is needed to assess the durability of merger signatures in streams.

This work aims at quantifying the dynamical impact of a galactic merger on a pre-existing population of stellar streams, with a particular focus on the emergence of asymmetries between the leading and trailing arms. We conduct N-body simulations of a Milky Way–like galaxy, initially hosting a population of 36 stellar streams, merging with a satellite galaxy ten times less massive than the host. 

In Section \ref{Numerical_methods}, we describe the simulation setup and the custom algorithm developed to extract stream density profiles. In Section \ref{Section_length_asymmetry}, we quantify the evolution of length asymmetries between the leading and trailing arms of the stream population throughout the simulation. In Section \ref{New asymmetry}, we motivate the development of a new method to characterize stream asymmetries and apply it to assess the impact of the merger more robustly.

\section{Numerical methods} \label{Numerical_methods}

\subsection{Simulation setup}

In order to investigate the formation and evolution of asymmetries in stellar streams, we perform a N-body simulation that models the accretion of a satellite galaxy onto a Milky Way-type galaxy, hosting a population of stellar streams. 

The host galaxy is represented by a multicomponent self-consistent galaxy model, which creates a Milky Way-like model fitted to Gaia DR2 kinematic data \citep{Binney2023}. The N-body initial conditions are generated using the AGAMA software suite \citep{Vasiliev2019}. Its total mass is $1.03\times10^{12}\text{M}_{\odot}$, and the parameters of its various components are summarized in Table \ref{tab:parameters_MW}. All components, including the gas, are represented by particles but the hydrodynamics equations are not solved. A Galatic bar naturally develops at $t=2\,\mathrm{Gyr}$.

\begin{table}
    \resizebox{\columnwidth}{!}{%
    \begin{tabular}{lcccc}
        \hline
        \hline
           & Mass [$10^{10}$Msun] & Scale radius [kpc] & Scale height [kpc] & Truncation radius [kpc] \\
        \hline
Thin disk    & 1.96 & 2.5  & -0.18 &       \\
Thick disk   & 0.83 & 2.1  & -0.7  &       \\
Gas disk     & 1.04 & 5.0  & -0.06 & 5.0   \\
Stellar halo & 0.10 & 30.7 &       & 11.5  \\
Bulge        & 1.23 & 1.0  &       & 0.7   \\
Dark halo    & 95.2 & 5.0  &       & 300.0 \\
        \hline
    \end{tabular}%
    }
    \caption{Parameters of the different components of the Milky Way-like galaxy.}
    \label{tab:parameters_MW}
\end{table}

In addition, the simulation includes 36 compact stellar objects, similar to globular clusters (GCs). These GCs contribute to the global potential and generate stellar streams. Their intrinsic properties (mass and size) and orbital parameters are selected from the e-TidalGCs catalogue \citep{Ferrone2023}, based on Milky-Way globular clusters \citep{Baumgardt2018, Baumgardt2021, Vasiliev&Baumgardt2021}. Selection criteria required the clusters to have sufficiently high energy ($>-10^{5}\text{km}^2/\text{s}^2$) to avoid full dissolution before the satellite's arrival. Each GC is modeled using a Plummer profile and a fixed particle mass of 6.5 $\text{M}_\odot$ is used. Consequently, the number of particles per GC is determined by dividing the total GC mass by this fixed mass. Although we adopt the parameters of real GCs, we stress that we do not aim at reproducing observed systems, but rather setup a realistic model for our numerical experiment.

We perform two simulations with this setup. A control simulation, hereafter referred to as the \texttt{reference case}, includes includes the host galaxy and the 36 GCs but no companion galaxy. The \texttt{merger case} adds the infalling satellite galaxy described below. The control simulation thus provides a basis for direct comparison to isolate the impact of the merger event.

The galaxy is modeled using a total of $2\times10^6$ particles: $1.5\times10^6$ for the dark matter halo and $5\times10^5$ for the baryonic components, which include the bulge, disks, and stellar halo. To balance accuracy and computational cost, we use galaxy particles with masses of $6.5\times10^5\text{M}\odot$ (dark matter) and $1.0\times10^5\text{M}\odot$ (baryons), while stream particles have much higher resolution at $6.5\text{M}\odot$. This choice ensures that individual galaxy particles remain below the typical mass range of $10^6-10^8\text{M}\odot$ of dark matter subhaloes known to create gaps in stellar streams \citep{Erkal2016}. Using more massive particles would increase the risk of mimicking such perturbers. With a mass of $6.5\times10^5\text{M}\odot$ and a softening length of 1\,pc, our dark-matter particles have sizes comparable to globular clusters and could produce gaps in the stellar streams \citep{Ferrone2025}. The adopted softening length of 1\,pc offers a compromise between reducing the graininess of the potential and preserving the physical realism of the internal structure of globular clusters. While the same particle sampling is used in both the reference and merger runs, ensuring that relative differences primarily trace the merger's impact, a population of low-mass, point-like particles can still induce artificial heating that accelerates stream phase-mixing \citep[e.g.,][]{Nibauer2025}. This effect operates similarly in both runs and may shorten the duration of the merger-induced asymmetry enhancement. For the same computational reasons, the simulation is strictly collision‑less: star-by-star relaxation is not modeled (see Section \ref{discussion_collisionless} for a full discussion of this approximation).

To simulate a galactic merger event, we include a satellite galaxy modeled as a scaled-down replica of the host galaxy for simplicity. Following the approach described by \citet{Pagnini2023}, the satellite mass is reduced by a factor of 10 and its size by a factor of $\sqrt{10}$, thereby preserving the same surface density as the primary system. The number of particles used to represent the satellite is also reduced by a factor of 10 compared to the host galaxy, ensuring consistent mass resolution. This mass ratio is consistent with estimates of the last major merger events in the Milky Way history, such as the Gaia-Sausage-Enceladus accretion \citep{Helmi2018, Belokurov2018}. Our goal is not to model a specific historical event, but to explore how a merger affects stellar streams in a realistic configuration.

The initial conditions for the main galaxy, the satellite and the globular clusters are specified in Cartesian coordinates, with the main galaxy centered at the origin and its disk aligned with the $x$–$y$ plane. The satellite is initialized at a distance of 250\,kpc along the $x$-axis, with an initial velocity vector that induces a prograde orbit relative to the host disk. To avoid alignment between the planes of the two disks, the satellite disk's orientation is adjusted through a rotation of 170° about the $y$-axis and 22° about the $z$-axis. 

The simulation is evolved using the gyrfalcON integrator \citep{Dehnen2002} from the NEMO stellar dynamics toolbox \citep{NEMO}. We adopt an opening angle parameter of $k_{\rm max} = 8$ and the same softening length of $\varepsilon = 1$\,pc for all the particles. These parameters were selected based on energy conservation and stability tests of individual globular clusters and the galaxies to ensure numerical robustness. The simulation is evolved for a total time of 10\,Gyr.

\begin{figure*}
\centering
  \includegraphics[width=17cm]{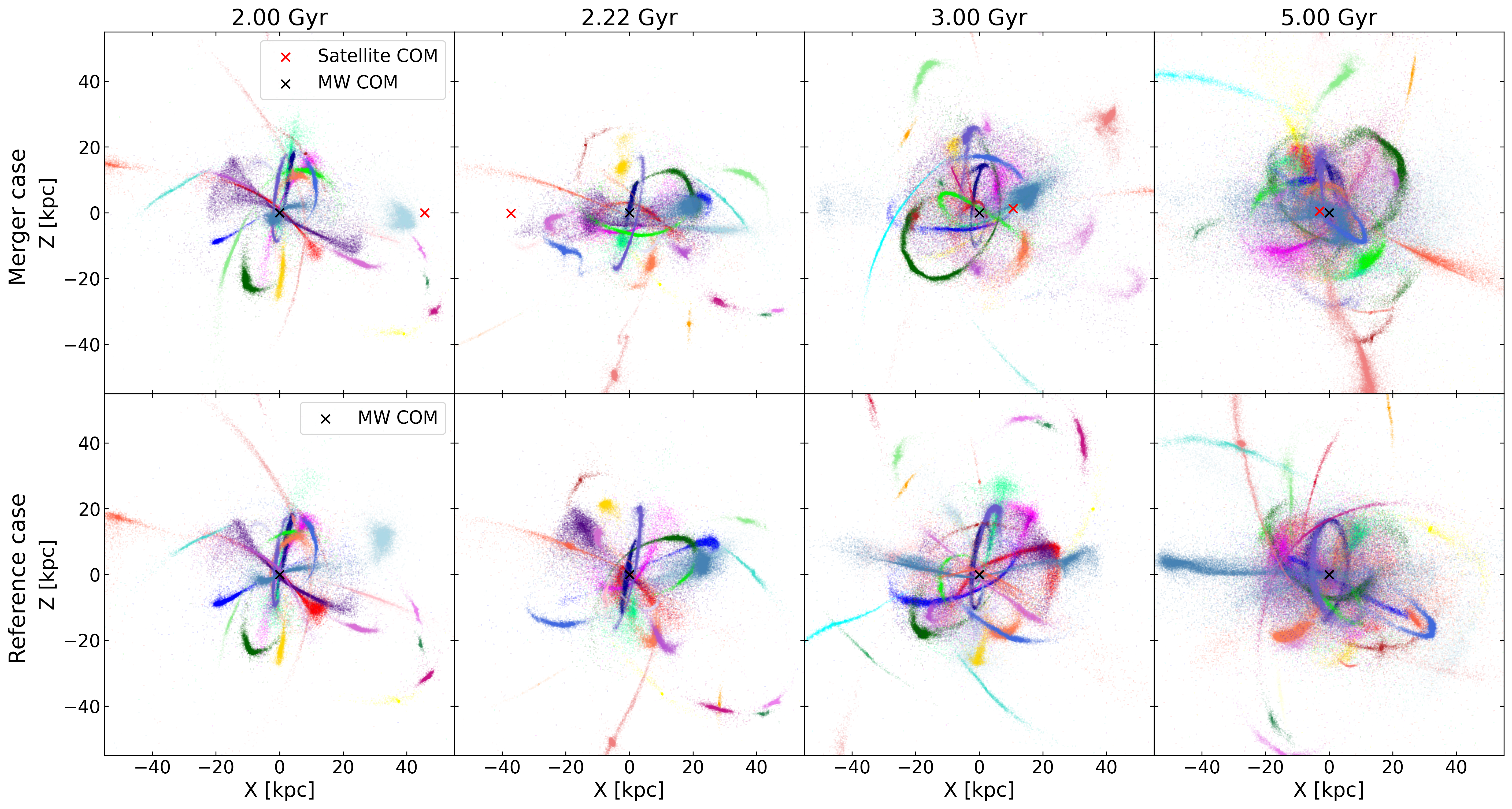} 
  \caption{Spatial distribution of stellar streams in the merger case (top row) and reference case (bottom row) at four different timesteps: 2.00\,Gyr (before the merger), 2.22\,Gyr (shortly after the satellite's first crossing of the Milky Way disk), 3.00\,Gyr, and 5.00\,Gyr. Each colour represents a different stellar stream. The black cross marks the centre of mass (COM) of the Milky Way-type galaxy, and the red cross marks the COM of the satellite galaxy (only in the merger case). All panels show projections in the $x-z$ plane, with coordinates centered on the Milky Way COM.}
  \label{Fig:simu}
\end{figure*}

The satellite crosses the host galaxy’s disk at 2.1\,Gyr and again at 2.8\,Gyr, before fully merging. Fig.~\ref{Fig:simu} provides a visual overview of the stream system at key stages before and after these crossings, illustrating both the reference and merger cases.

\subsection{Stream Detection and Analysis with 1-DREAM} \label{1-dream}

In previous studies, stream density profiles have commonly been derived by projecting particles along the trajectory of the progenitor globular cluster \citep[e.g.,][]{Kupper2010}. This method assumes that the stream closely follows the orbit of its progenitor, allowing for the construction of a curvilinear coordinate along which the particle distribution can be measured. Alternatively, when streams are relatively short, a stream-aligned coordinate system can be defined to directly extract the density profile - as is commonly done for observational studies of streams such as GD-1 \citep[e.g.,][]{Ibata2020}. However, these techniques become impractical for streams that are both long and significantly wrapped around the galaxy, such as those formed in our simulations. The complex morphology of such streams makes it impossible to define a single trajectory or coordinate system suitable for accurate projection and profiling. This necessitates a more flexible and automated approach. 

To this end, we utilize the 1-DREAM framework \citep[1D Recovery, Extraction, and Analysis of Manifolds in noisy environments,][]{1DREAM}. This machine learning-based method is designed to extract filamentary structures embedded in noisy point clouds and is well suited to the detection of dynamically evolved stellar streams. In practice, 1-DREAM uses an ant-colony–style pheromone field to detect dense one-dimensional structures from a 3D distribution of particles. After temporarily filtering the lowest densities and adjusting the distribution of points according to the pheromones, a local eigen-decomposition is used to identify filamentary structures, which are then followed by are crawler to connect them into a skeleton. Before applying 1-DREAM, we wait for sufficient dynamical evolution to ensure the stream is formed. To improve the computational efficiency of the algorithm, we remove the cluster itself by identifying the particle with the densest environment in the cluster (found using a k-d tree density estimate with five nearest neighbours) and excluding all particles within a radius of 1\,kpc from this point. This step accelerates the convergence of the filament detection by eliminating the dense central clump. We then apply the 1-DREAM methodology using its stellar stream detection workflow. This process can identify multiple filamentary structures, but we retain only the longest filament passing through the cluster’s original location. This ensures we focus on the principal stream, minimizing the influence of spurious detections or fragmented substructures. Once the filament is detected, we resample it uniformly, defining bins of 1\,kpc along its length. This binning regularizes the filament’s spatial structure and allows consistent comparisons across streams. To construct the density profile along the stream, we first assign particles detected by 1-DREAM to the nearest bin along the resampled filament and reintroduce particles initially excluded from the cluster by associating them with their closest bin. We then count the number of particles in each bin to obtain the density profile. 1-DREAM extracts the filament without assigning a preferential direction, so we do not initially know which side corresponds to the leading or trailing arm. By computing the dot product between the mean velocity vector of particles in each bin and the local tangent to the filament, we detect and correct any inversion, allowing us to separate both arms.

\begin{figure*}
\centering
  \includegraphics[width=17cm]{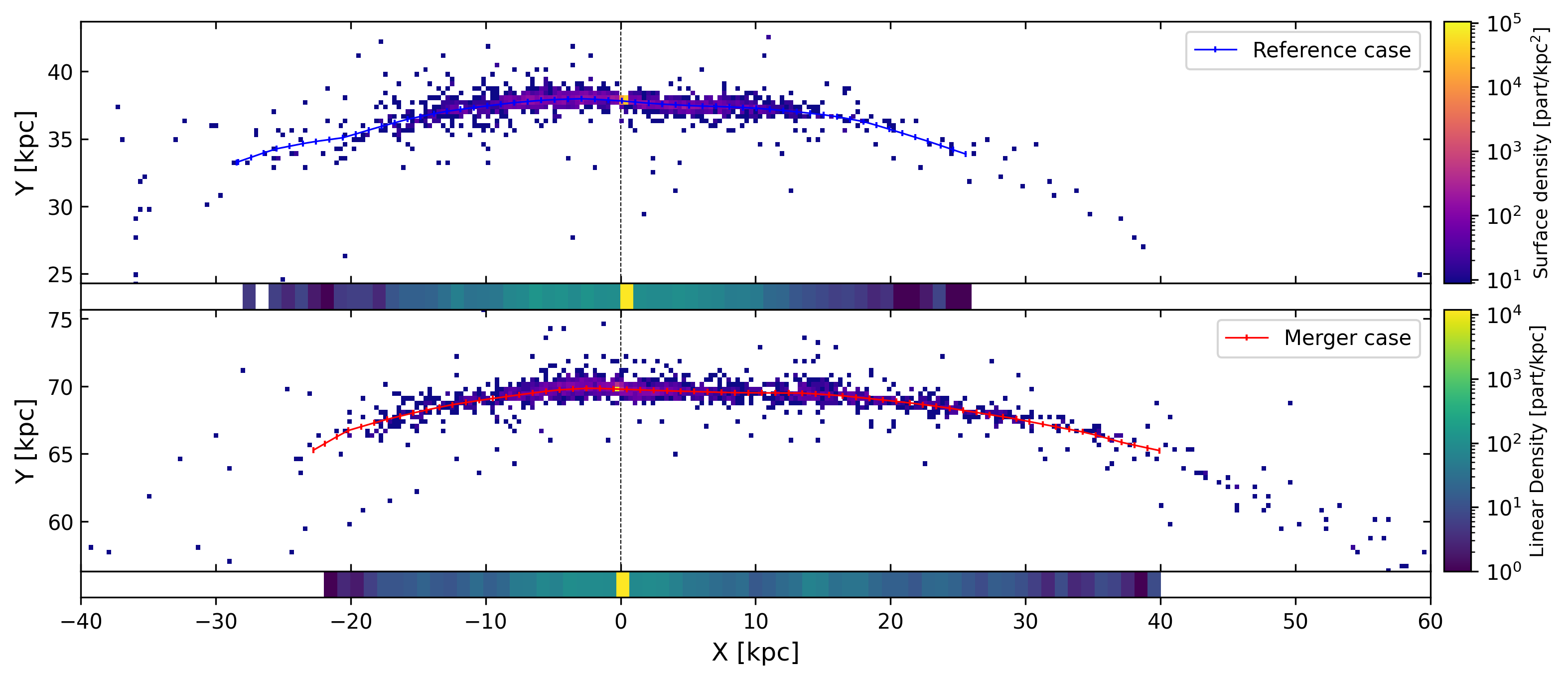} 
  \caption{Projection of a stellar stream from our simulations in its progenitor’s orbital plane at $t=3$\,Gyr, for the reference run (top) and the merger run (bottom). The Milky Way’s centre of mass is at $(x,y)=(0,0)$ and the progenitor is located at x=0. The solid curves are the 1-DREAM fits to the stream (blue for the reference case, red for the merger case). The panels below each map display the corresponding density profiles along the stream.}
  \label{Fig:Ter8_maps}
\end{figure*}

To illustrate this procedure, we extract one stellar stream from both the reference and merger simulations. A snapshot of this stream at $t = 3$\,Gyr is shown in Fig.~\ref{Fig:Ter8_maps}. The coordinates are centered on the progenitor, located at $x=0$ kpc, allowing a direct visual comparison between the two cases. The over-plotted curve shows the 1-DREAM filament fit (only one filament is detected in both cases), while the density profile along the filament is displayed below each stream map. The time evolution of this density profile is shown separately in Fig.~\ref{Ter8_time_evolution}. In the merger case, the leading arm (positive $x$) extends farther than in the reference case, and an under-density is visible at $x \approx 10$ kpc.

Visual inspection of the merger simulations reveals strongly distorted stream morphologies, including gaps, folds, bifurcations, and even stellar ejections. Our density–profile method with \textsc{1-DREAM} cannot capture all such features, particularly folds and bifurcations. However, since these detection limitations affect both the reference and merger runs equally, a direct comparison still isolates the merger’s specific impact.

\section{Length asymmetry} \label{Section_length_asymmetry}

Length asymmetries between the leading and trailing arms of a stream can arise from both internal dynamics and external perturbations. In modified gravity frameworks such as MOND, asymmetries emerge through the external field effect: particles tend to escape the progenitor in a preferential direction, thereby generating strongly asymmetric tidal tails, as shown by \citet{Thomas2018} for a Palomar 5–like stream. In addition, eccentric orbits naturally produce alternating stretching and contraction of the arms, leading to length differences between the two arms. Furthermore, external perturbations such as torques from a rotating Galactic bar can truncate one arm, reproducing the observed Palomar 5 signature in which the trailing arm extends nearly twice as far as the leading arm \citep{Pearson2017}. In this section, we aim to quantify length asymmetries across our simulated streams to test whether galactic mergers can generate signatures of comparable magnitude.

\subsection{Length asymmetry caused by eccentric orbits}

\begin{figure}  
  \resizebox{\hsize}{!}{\includegraphics{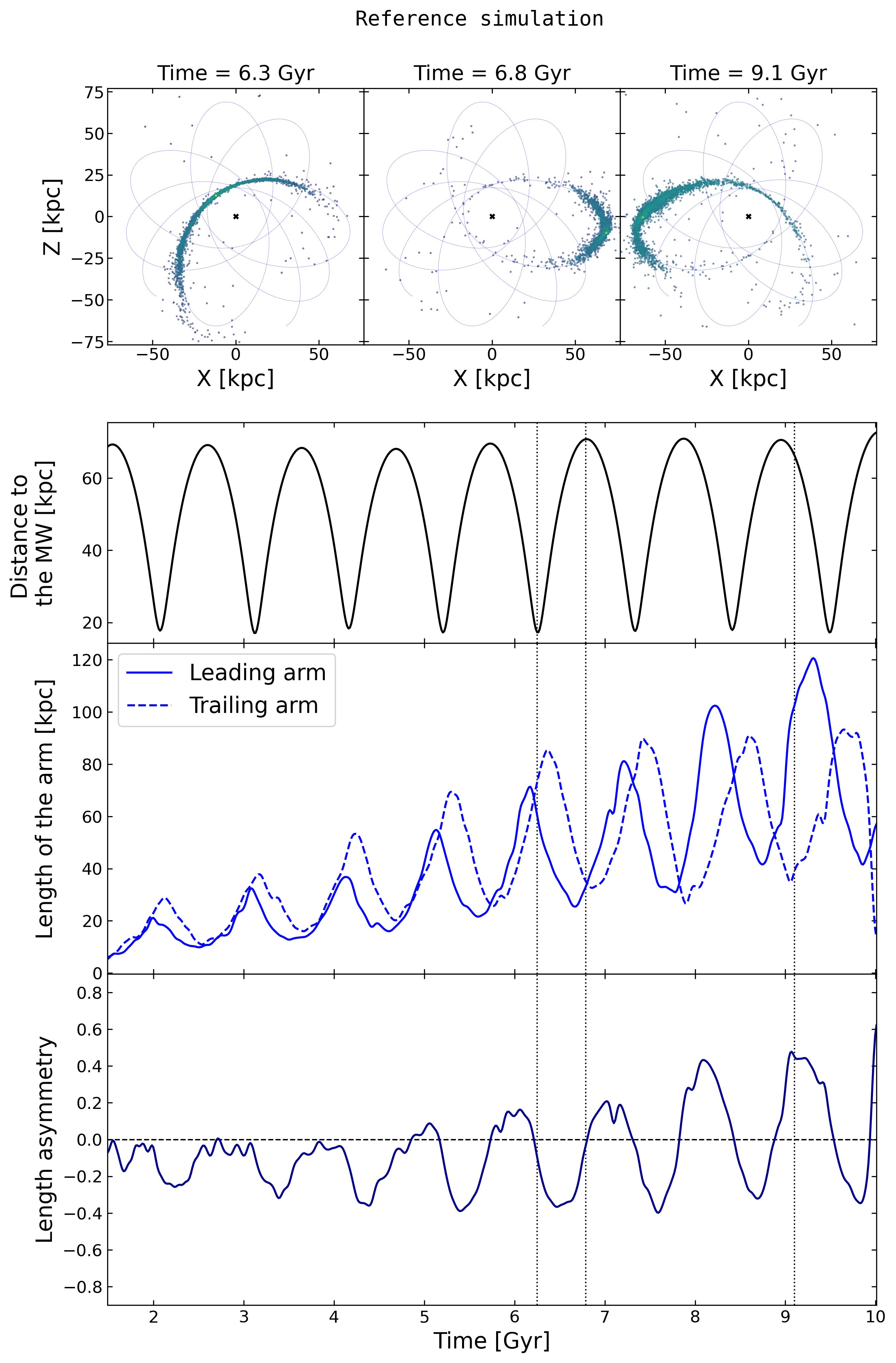}}
  \caption{\textit{First row:} Three snapshots of one stream in the reference simulation, shown at the progenitor’s pericentre, apocentre, and at a time when the length asymmetry reaches one of its maxima. The black cross marks the Milky Way’s centre of mass and the blue line the progenitor’s trajectory. \textit{Second row:} Time evolution of the progenitor’s distance to the Milky Way’s centre of mass. \textit{Third row:} Time evolution of the length of the leading and trailing arm. \textit{Fourth row:} Normalized asymmetry between the leading and trailing arm lengths. The three vertical dotted lines mark the times of the snapshots shown in the top row. All time-dependent curves are smoothed with a 1D Gaussian filter with $\sigma=2$.}
  \label{Ter8_length_asymmetry}
\end{figure}

When measuring the asymmetry of stellar stream arm lengths, we must account for the fact that the stream alternately stretches towards pericentre and contracts towards apocentre. As the stream grows longer, the arms are tidally stretched at different times, naturally producing an asymmetry along any eccentric orbit. To illustrate this effect, a single stream is extracted and analysed from the reference simulation (the same stream shown in the upper panel of Fig.~\ref{Fig:Ter8_maps}). Additional properties of this stream are examined in more detail in Section \ref{Single stream analysis}. Fig.~\ref{Ter8_length_asymmetry} shows the evolution of the lengths of the leading arm and the trailing arm, and their normalized asymmetry
\begin{equation}
    A_{length} = \frac{l_l - l_t}{l_l + l_t},
\end{equation}
where $l_l$ is the length of the leading arm and $l_t$ is the length of the trailing arm, measured using the method described in Section \ref{1-dream}. When the stream is short, both arms pass through the pericentre with a short time difference. Therefore, the asymmetry measurement fluctuates with a small amplitude around zero. As the stream lengthens, however, the time interval between the leading and trailing ends of the stream passing through pericentre increases: the leading arm stretches significantly before the trailing arm, and the amplitude of the asymmetry oscillations grows markedly.

When studying the asymmetry induced by an external process, a galactic merger in this study, we must account for the streams' asymmetry caused by their eccentric orbits.

\subsection{Statistics over the population of stellar streams}

\begin{figure}  
  \resizebox{\hsize}{!}{\includegraphics{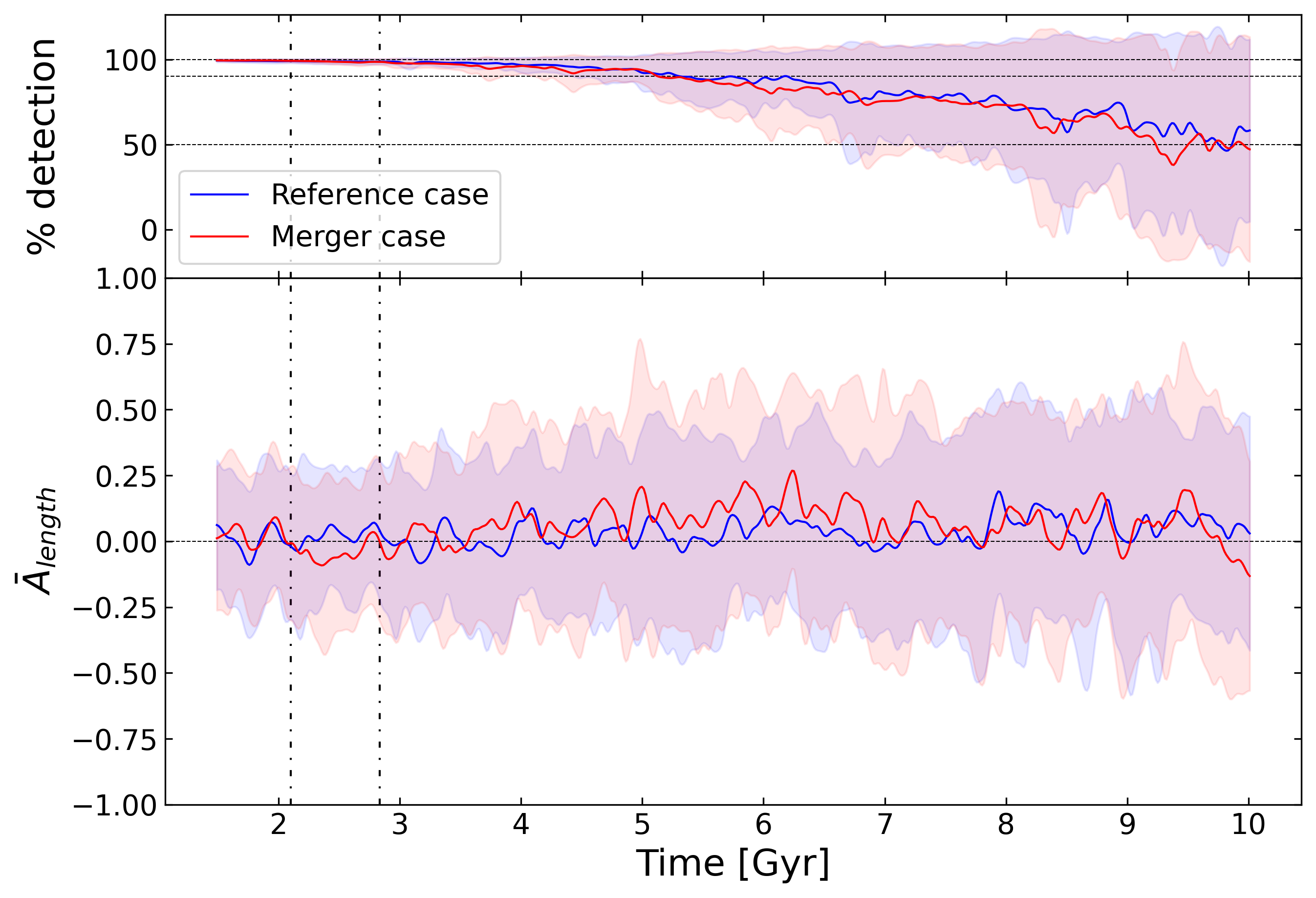}}
  \caption{\textit{Top row:} Fraction of stream particles recovered by the 1-DREAM algorithm as a function of time. \textit{Bottom row:} Median asymmetry between the leading and trailing arm lengths versus time; shaded envelopes indicate the $\pm1\sigma$ robust standard deviation. The dash-dotted lines mark the first (2.1\,Gyr) and second (2.8\,Gyr) passages of the satellite.}
  \label{Length_asymmetry}
\end{figure}

At each timestep, we measure this length asymmetry for the 36 streams and show in Fig.~\ref{Length_asymmetry} the evolution of the median over the population. As we showed above, the intrinsic length asymmetry of a single stream fluctuates, with increasing amplitude as the stream lengthens. In our setup, all 36 streams start forming simultaneously. Therefore, their length asymmetries all share fluctuations of small amplitude. Because each stream follows a different orbit, their asymmetry fluctuations are not synchronized, keeping the population median close to zero. As time progresses and streams grow longer, the dispersion of the reference case increases, indicating larger asymmetry amplitudes. However, the absence of synchronization keeps the median small. In the merger run, the satellite first crosses the host’s disk at 2.1\,Gyr and again at 2.8\,Gyr, before fully merging. The dispersion in the length asymmetry increases soon after the merger around $t\approx3.5$\,Gyr, and the median asymmetry shifts to positive values from 3.5 to 7\,Gyr.

The merger-induced asymmetry signal remains modest for two main reasons. First, our length measurements rely on 1-DREAM, which struggles to recover streams that are strongly wrapped or disrupted near the MW: these streams undergo stronger tidal stripping, fall below detection limits, which causes the decrease of particle counts starting at $t\approx5$\,Gyr seen in the upper panel. Second, most streams already exhibit length asymmetry caused by their eccentric orbit, so any additional distortion from the merger is diluted by averaging over unsynchronized fluctuations.

In Appendix \ref{Annexe_3_populations}, we classify stellar streams into three groups based on the timing of their closest interaction with the satellite galaxy. We find that the streams affected before and after the merger, i.e. those on wide orbits around the Milky Way, remain detectable and exhibit modest length asymmetries that persist over long timescales. This suggests that, to uncover asymmetries imprinted by a past galactic merger, one should target streams on wide orbits, where the weaker tidal field permits their prolonged survival.

\section{Accounting for density features in tail asymmetries} \label{New asymmetry}

\subsection{The need for a new asymmetry measurement} \label{need_new_asym}

A merger induces structural complexity, producing both under- and over-densities. The previous length measurement does not fully capture this complexity. To catch these asymmetries in density, we can measure the number of particles in a fixed range $d_1-d_2$ from the progenitor, and compute the ratio of the two arms:
\begin{equation}
    q=\frac{N_{l,d_1-d_2}}{N_{t,d_1-d_2}},
\end{equation}
 where $N_{l,d_1-d_2}$ (respectively $N_{t,d_1-d_2}$) is the number of stars in the leading arm (resp. trailing) in the range $d_1-d_2$ from the progenitor. \citet{Kroupa2022} uses the $q$-parameter over the range $50-200$\,pc from the progenitor to measure asymmetries in open clusters.
 
 While applying this technique to our sample of stellar streams, no clear trend emerges that could illustrate the effect of the merger (see Appendix \ref{Annexe_asymmetry_nb_particules+density} for details). This likely comes from the fact that these methods were originally designed to capture the external field effect in MOND, which acts continuously and uniformly on stellar streams \citep{Thomas2018}. In contrast, the perturbation induced by the merger in our simulations is external and localized, with its impact varying from stream to stream. Moreover, it is important to note that studies employing these methods within the MOND framework typically consider clusters on near-circular orbits. In contrast, the clusters in our simulations follow significantly more eccentric trajectories. As a result, this asymmetry diagnostic is not well suited to capture the complex structural disturbances caused by a galactic merger. What is needed is an approach that is sensitive to the structural differences between the leading and trailing arms, independent of their position along the stream, and that can accumulate these differences into a single robust measurement. 

To address this, we propose a new method to quantify stream asymmetry based on comparing the cumulative stellar density profiles of the leading and trailing arms. Each cumulative distribution function is defined as
\begin{equation} \label{eq_cumulative_function}
F(k) = \frac{1}{N}\sum_{j=2}^{k} n_j,
\end{equation}
where $n_j$ is the number of particles in bin $j$, $k$ is the number of bins along the arm,  and $N$ is the total number of particles in the arm. We start at 2\,kpc outward from the progenitor to avoid any contamination from the progenitor itself.

Since the two arms may differ in length and bin number, we first normalize their cumulative distribution functions and then determine their overlapping domain. In this section, $D$ represents the distance along one tidal arm spine, starting at 2\,kpc from the progenitor, while $D_{tot}$ is the total arm length determined by 1-DREAM described in Sect.~\ref{1-dream}. A monotonically increasing common grid $\{x_i\}_{i=1}^{M}$ is constructed, with $M$ the number of grid points (here we adopt $M=100$). The normalized cumulative profiles are linearly interpolated onto this grid. The asymmetry is then defined as the sum of the absolute difference between the two profiles over their common domain:
\begin{equation}\label{eq:asym}
A = \sum_{k=1}^M\big|F_{\mathrm{lead}}(k)-F_{\mathrm{trail}}(k)\big|.
\end{equation}
Over- or under-densities or structural disturbances make the cumulative functions diverge, and the asymmetry increases accordingly.

\begin{figure}  
  \resizebox{\hsize}{!}{\includegraphics{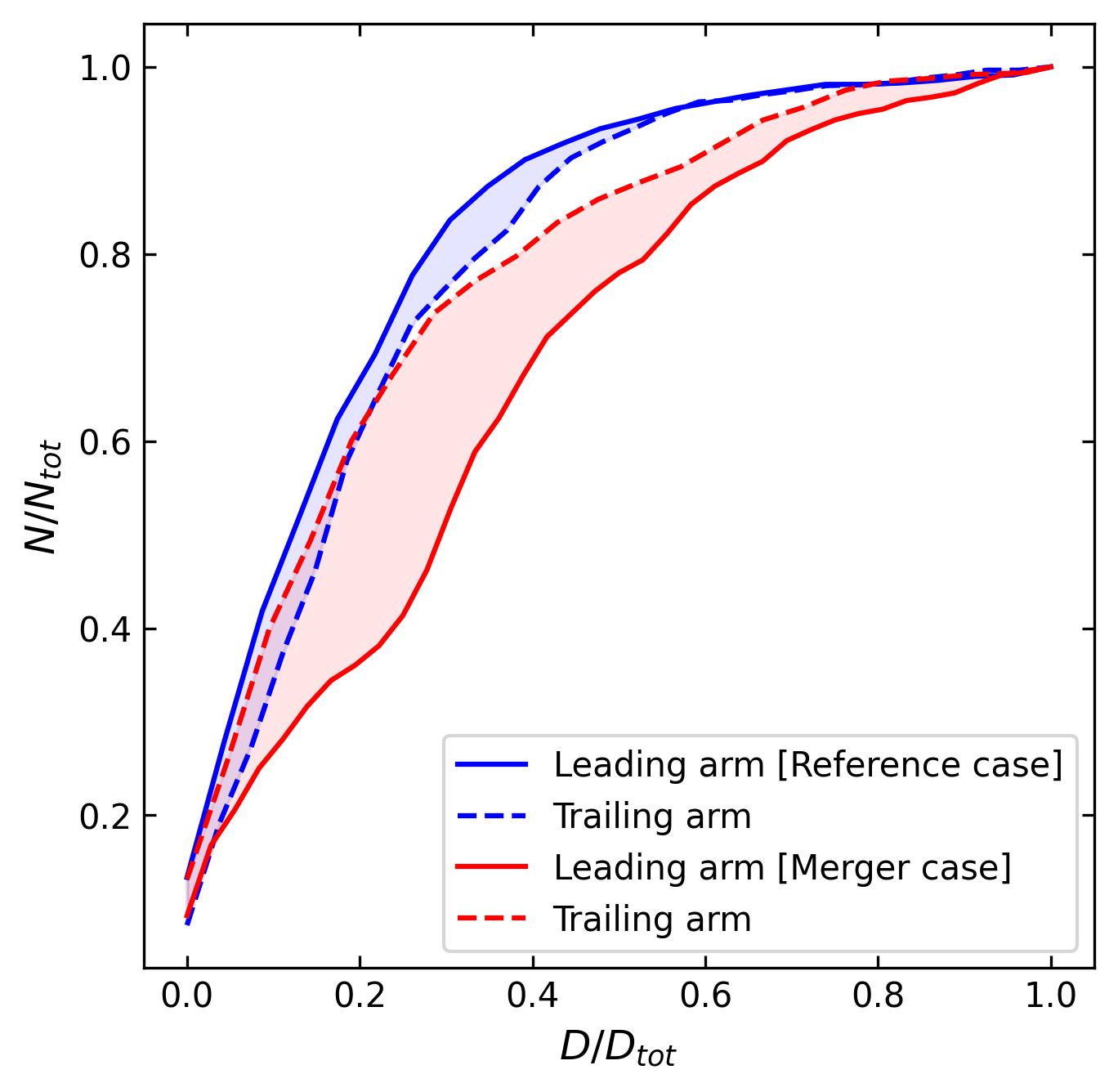}}
  \caption{Normalized cumulative fraction of particles along the leading (solid) and trailing (dashed) arms of the same stream shown in Fig.~\ref{Fig:Ter8_maps}, computed from 2\,kpc outward from the progenitor. Blue curves correspond to the reference simulation and red curves to the merger simulation.}
  \label{fig:cumulative_function}
\end{figure}

Fig.~\ref{fig:cumulative_function} shows these cumulative functions for the stellar stream presented in Fig.~\ref{Fig:Ter8_maps} at $t=3$\,Gyr. In the reference case (blue curves), the two cumulative curves nearly coincide, resulting in a small shaded area between them, which corresponds to a low asymmetry A value. In contrast, in the merger case, the under-density at 10\,kpc creates a plateau at $D/D_{tot}=0.2$ in the cumulative curve of the leading arm (solid red curve). This causes a larger separation between the two red curves, increasing the shaded area and thus the measured asymmetry.

\subsection{Single stream analysis} \label{Single stream analysis}

\begin{figure*}
\centering
   \includegraphics[width=17cm]{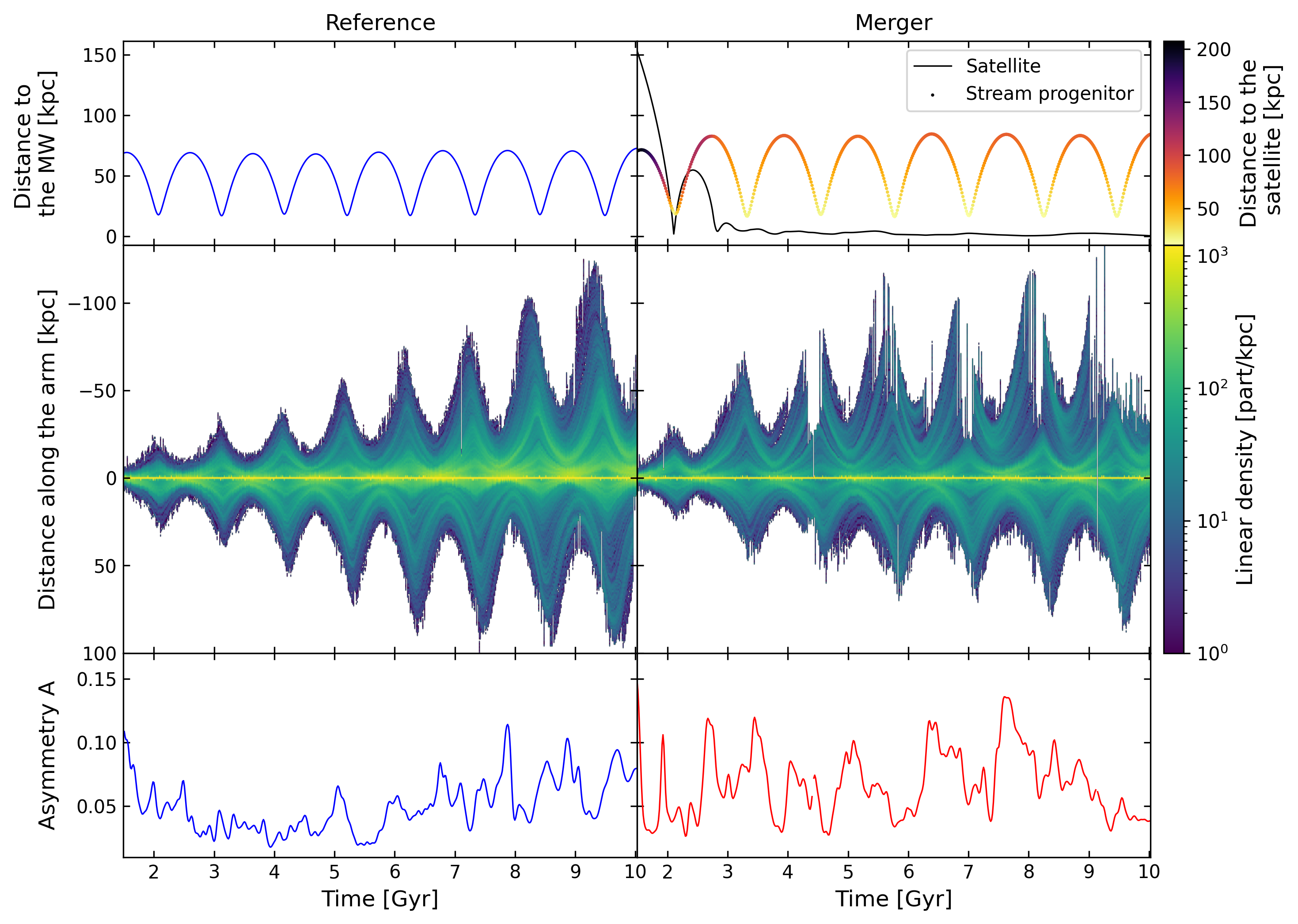}
     \caption{Time evolution of a single stellar stream from our simulations. The left column shows the reference run and the right column shows the merger run. \textit{Top panel:} Distance of the stream progenitor to the Milky Way’s centre of mass as a function of time. In the merger case, each point is colour‐coded by the progenitor’s distance to the satellite, and the satellite’s own orbital radius around the Milky Way is overplotted in black. \textit{Middle panel:} Evolution of the density profile along the stream (zero at the progenitor; negative values denote the leading arm, positive values the trailing arm). \textit{Bottom panel:} Evolution of the stream asymmetry, defined as the sum of the absolute difference between the two cumulative density profiles of the leading and trailing arms (Eq.\ref{eq:asym}).}
     \label{Ter8_time_evolution}
\end{figure*}

We track this stream’s asymmetry over the full simulation in Fig.~\ref{Ter8_time_evolution}. The top row shows the progenitor’s distance to the Milky Way’s centre; in the merger case, we also plot the satellite’s distance to the MW in black and colour-code the progenitor curve by its separation from the satellite. The middle row displays the stream’s density profile over time, and the bottom row shows the asymmetry.

Notably, the merger takes place when the stream is at pericentre ($t = 2$\,Gyr), after which the stream shifts to a higher orbit, in line with the conclusion of \citet{Weerasooriya25}. As expected, the density profiles show repeated stretching at pericentres and contraction at apocentres. In the reference case, alternating under- and over-densities appear along the stream, caused probably by the epicyclic motion of stars in the stream’s own reference frame \citep{Kupper2008, Kupper2010, Just2009}. The stream lengthens steadily over time, and the measured asymmetry remains low throughout 2–6\,Gyr.

In the merger case, the same epicyclic pattern is still visible, but additional under-densities emerge in the leading arm (negative coordinates) after the merger. As these features are stretched through pericentre, their densities can fall below 1-DREAM’s detection threshold. Unlike the gradual growth of the reference stream, the merger stream undergoes a sudden extension immediately after the merger, followed by slower growth. Correspondingly, the amplitude of the asymmetry increases at the time of the merger, driven by the formation of gaps in the leading arm that create a clear structural imbalance between the two arms. At the time of the satellite's first passage, the leading arm is located at pericentre. The particles within this arm experience a deeper potential, resulting in greater acceleration compared to the trailing particles which are not impacted. This differential acceleration leads to the formation of gaps in the leading arm.

\subsection{Statistics over the population of stellar streams}
 
\begin{figure}
  \resizebox{\hsize}{!}{\includegraphics{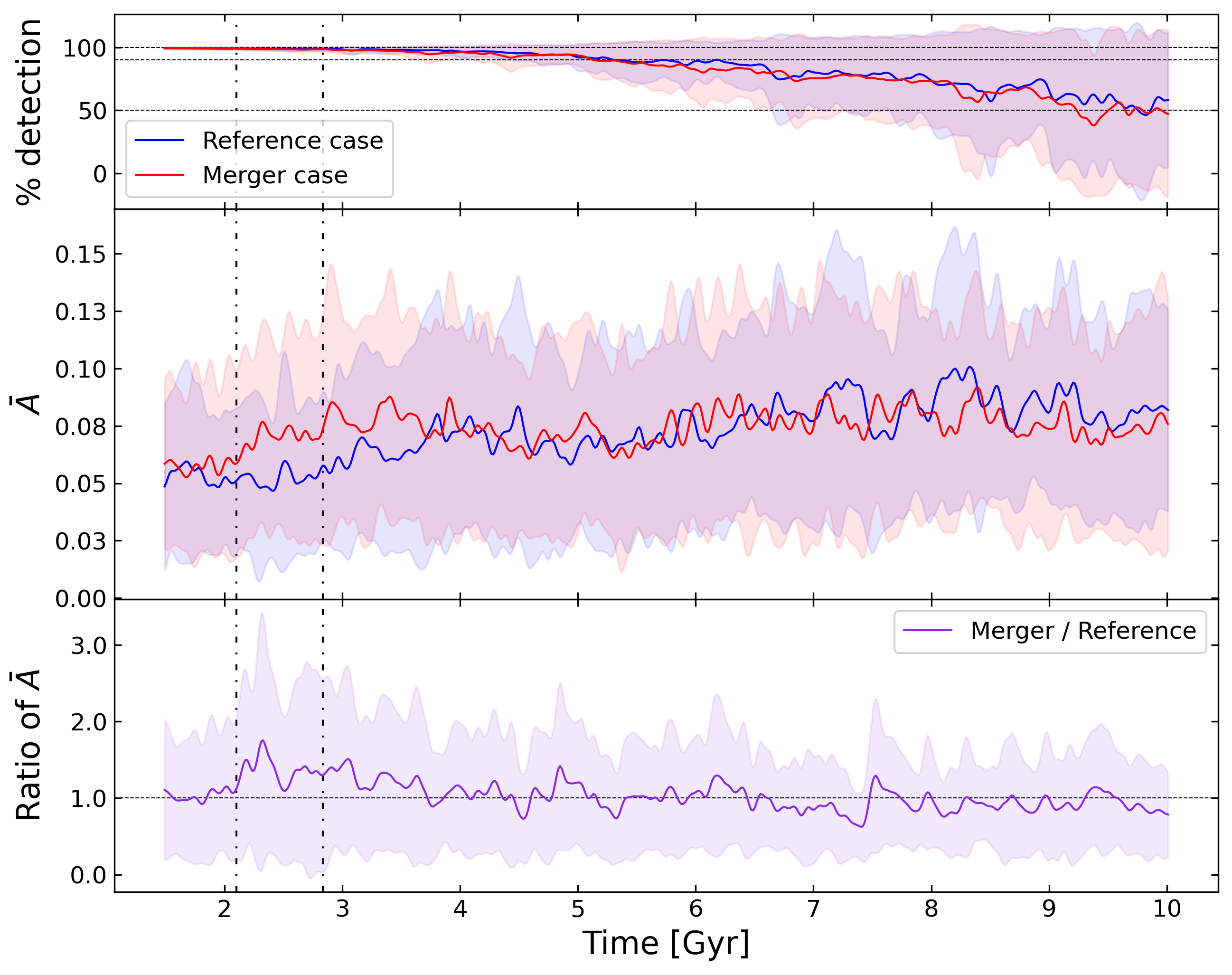}}
  \caption{\textit{Top panel:} Fraction of stream particles recovered by the 1-DREAM algorithm as a function of time. \textit{Middle row:} Median asymmetry between the leading and trailing arms versus time. \textit{Bottom panel:} Ratio of the median asymmetry in the merger case to that in the reference case. Shaded envelopes indicate the $\pm1\sigma$ robust standard deviation.}
  \label{New_asymmetry}
\end{figure}

Repeating this method for all 36 streams, we compute the median asymmetry across the entire population at each time step and track its evolution throughout the simulation, as shown in the middle row of Fig.~\ref{New_asymmetry}. As in Fig.~\ref{Length_asymmetry}, the top row displays the median fraction of stream particles detected by the 1-DREAM algorithm, also calculated over the 36 streams at each time, with shaded area indicating the $\pm1\sigma$ dispersion. The merger at $2\,\mathrm{Gyr}$ enhances asymmetry, with effects lasting for about $2\,\mathrm{Gyr}$. Over time, the particles drift farther from the globular cluster, and the resulting gaps and over-densities move outward. The bottom row further illustrates this by showing the ratio of merger to reference asymmetry, highlighting that the median enhancement remains above unity for about 2\,Gyr after the merger, with the shaded band representing the scatter across the population of streams. As they spread out, these features become progressively more diffuse, eventually slipping below the detection threshold of the 1-DREAM method, as reflected in the declining detection fraction. By then, the only stars detected as part of the stream have left the progenitor after its encounter with the satellite. Therefore the section of the stream made of these stars does not reflect the perturbation from the merger event.

\section{Discussion}

\subsection{Limitations of the collision-less approximation} \label{discussion_collisionless}

A methodological caveat of this study is that our simulations are strictly collision‑less: particles evolve only in a smooth gravitational potential. This suppresses close encounters and the two‑body relaxation that, in reality, shapes the cluster’s internal evolution and modulates its star‑loss rate \citep{Spitzer1987}. In realistic clusters, stars escape if their energy exceeds the threshold set by the Lagrange points $L_1$ and $L_2$ and if they subsequently pass through the vicinity of these points, otherwise, they may remain bound for extended periods \citep[see, e.g.][]{FukushigeHeggie2000}. Relaxational encounters redistribute energy, allowing some stars to reach this threshold more rapidly, while others sink toward the centre (mass segregation). In dense cores, repeated interactions form and harden binaries that act as an internal energy source and halt core collapse \citep{Davies2013}. Stellar evolution further modifies the energy balance and binary dynamics, enhancing mass loss and escape \citep[see, e.g.][]{Joshi2001}. Accurately modelling these effects requires calculating the gravitational interactions between every pair of stars, which scales as $\mathcal{O}(N^2)$. This makes fully collisional simulations computationally too expensive for galaxy-scale merger studies.

However, once stars are unbound, the stellar‐density in the tidal tails is so low that two body gravitational encounters are negligible. Indeed, \citet{Mastrobuono-Battisti2012} demonstrate that stream geometry, kinematics, and epicyclic substructure are reproduced equally well in purely collision‑less and full N‑body models. Thus, although the internal dynamics of our progenitors differ from fully collisional expectations, the collision‑less approximation affects both leading and trailing arms, and it is present in both the merger and reference runs. Measuring asymmetries relative to the reference therefore still isolates the additional impact of the satellite merger on the pre‑existing stream population.

\subsection{Addition of merger-induced asymmetries on top of known effects}

Even in the reference simulation, which lacks the infalling satellite, several streams display gaps, bifurcations, and fan-like structures. These may arise from disk crossings or interaction with the Galactic bar, but further work is needed to establish the precise causes. Another possible contribution comes from the globular clusters themselves: as shown by \citet{Ferrone2025}, clusters with typical masses of a few $10^5\mathrm{M}_\odot$ can produce gaps in stellar streams. However, because the bar, the disk, and GC-size particles are present in both simulations, any additional perturbation seen only in the merger run can be attributed to the merger itself.

Stellar streams are subject to a variety of environmental influences. Besides merger events, numerous studies have reported gaps produced by substructure in the Galactic potential like dark-matter sub-halos \citep[e.g.][]{Carlberg2011,Sanders2016,Yoon2011} and by encounters with giant molecular clouds, disk crossings, or bar passages \citep{Amorisco2016,Pearson2017}, and more recently by globular clusters themselves \citep{Ferrone2025}. Such processes often act asymmetrically along the stream. In the context of modified gravity theories, \citet{Thomas2018} further showed that similar asymmetries can arise due to the breaking of the strong equivalence principle, affecting the symmetry of the progenitor's gravitational field with a subtle dependence on the progenitor's mass and orbit.  

Perturbations caused by major mergers can in any case add to an already rich set of signatures, making the interpretation of stream properties degenerated and more challenging.

\section{Conclusion}

We quantify how a 1:10 galactic merger imprints asymmetries on a pre-existing population of stellar streams by comparing an N-body simulation of a Milky Way–like galaxy hosting 36 streams from globular cluster-mass objects. We also run an additional simulation without the galactic interaction, for reference. Using a manifold-extraction algorithm to detect the streams, we introduce a new asymmetry measurement based on the difference between the cumulative density profiles of the two arms. This metric allows us to detect local under/over-densities and structural disturbances regardless of their positions along the stream, whereas classical length-based measurements lack this capability. We find that the galaxy merger enhances arm asymmetries shortly after coalescence of the galaxies. As the streams continue to get elongated, the induced features slowly dilute far away from the progenitor. On average, asymmetric signatures disappear within $\approx 2$\,Gyr.

When averaging over the entire population of streams, the length asymmetry only shows a mild response to the merger, because of the non-simultaneity of their responses to the external stimulus, due to their various orbital parameters and relative distance to the perturber. The strongest, most persistent imprints occurs in streams with wide orbits that are perturbed prior to coalescence during a close passage of the infalling satellite, making them prime targets to uncover fossil merger signatures. Streams close to the main galaxy are affected as well but are rapidly scattered, making their asymmetries fade quickly.

Merger-driven asymmetries are of similar nature as those originating from other other known perturbations such as interactions with the galactic bar or spiral arms, disk shocking, and encounters with dwarf satellites, giant molecular clouds, and dark matter subhaloes. These processes produce similar observational signatures, creating degeneracies, such that an individual stream could rarely provide an unambiguous signature to decode its formation history. Consequently, asymmetric underdensities in stellar streams do not provide robust constraints on dark-matter subhaloes if a recent major merger is plausible.

\begin{acknowledgements}
We thanks the referee for a constructive report. CG and FR acknowledge support provided by the University of Strasbourg Institute for Advanced Study (USIAS), within the French national program Investment for the Future (Excellence Initiative) IdEx-Unistra. GFT acknowledges support from the Agencia Estatal de Investigaci\'on del Ministerio de Ciencia en Innovaci\'on (AEI-MICIN) and the European Regional Development Fund (ERDF) under grant number PID2020-118778GB-I00/10.13039/501100011033, and from the European Union Widening Participation ExGal-Twin (GA 101158446). G.P. acknowledges the support from the Centre national d'études spatiales (CNES) through a postdoctoral fellowship.
\end{acknowledgements}

\bibliographystyle{aa}  
\bibliography{biblio}

@ARTICLE{Binney2023,
       author = {{Binney}, James and {Vasiliev}, Eugene},
        title = "{Self-consistent models of our Galaxy}",
      journal = {\mnras},
         year = 2023,
        month = apr,
       volume = {520},
       number = {2},
        pages = {1832-1847},
          doi = {10.1093/mnras/stad094},
archivePrefix = {arXiv},
       eprint = {2206.03523},
 primaryClass = {astro-ph.GA},
       adsurl = {https://ui.adsabs.harvard.edu/abs/2023MNRAS.520.1832B}
}

@ARTICLE{Vasiliev2019,
       author = {{Vasiliev}, Eugene},
        title = "{AGAMA: action-based galaxy modelling architecture}",
      journal = {\mnras},
         year = 2019,
        month = jan,
       volume = {482},
       number = {2},
        pages = {1525-1544},
          doi = {10.1093/mnras/sty2672},
archivePrefix = {arXiv},
       eprint = {1802.08239},
 primaryClass = {astro-ph.GA},
       adsurl = {https://ui.adsabs.harvard.edu/abs/2019MNRAS.482.1525V}
}

@ARTICLE{Kruijssen2020,
       author = {{Kruijssen}, J.~M. Diederik and {Pfeffer}, Joel L. and {Chevance}, M{\'e}lanie and {Bonaca}, Ana and {Trujillo-Gomez}, Sebastian and {Bastian}, Nate and {Reina-Campos}, Marta and {Crain}, Robert A. and {Hughes}, Meghan E.},
        title = "{Kraken reveals itself - the merger history of the Milky Way reconstructed with the E-MOSAICS simulations}",
      journal = {\mnras},
         year = 2020,
        month = oct,
       volume = {498},
       number = {2},
        pages = {2472-2491},
          doi = {10.1093/mnras/staa2452},
archivePrefix = {arXiv},
       eprint = {2003.01119},
 primaryClass = {astro-ph.GA},
       adsurl = {https://ui.adsabs.harvard.edu/abs/2020MNRAS.498.2472K}
}

@ARTICLE{Helmi2020,
       author = {{Helmi}, Amina},
        title = "{Streams, Substructures, and the Early History of the Milky Way}",
      journal = {\araa},
         year = 2020,
        month = aug,
       volume = {58},
        pages = {205-256},
          doi = {10.1146/annurev-astro-032620-021917},
archivePrefix = {arXiv},
       eprint = {2002.04340},
 primaryClass = {astro-ph.GA},
       adsurl = {https://ui.adsabs.harvard.edu/abs/2020ARA&A..58..205H}
}

@ARTICLE{Malhan2022,
       author = {{Malhan}, Khyati and {Ibata}, Rodrigo A. and {Sharma}, Sanjib and {Famaey}, Benoit and {Bellazzini}, Michele and {Carlberg}, Raymond G. and {D'Souza}, Richard and {Yuan}, Zhen and {Martin}, Nicolas F. and {Thomas}, Guillaume F.},
        title = "{The Global Dynamical Atlas of the Milky Way Mergers: Constraints from Gaia EDR3-based Orbits of Globular Clusters, Stellar Streams, and Satellite Galaxies}",
      journal = {\apj},
         year = 2022,
        month = feb,
       volume = {926},
       number = {2},
          eid = {107},
        pages = {107},
          doi = {10.3847/1538-4357/ac4d2a},
archivePrefix = {arXiv},
       eprint = {2202.07660},
 primaryClass = {astro-ph.GA},
       adsurl = {https://ui.adsabs.harvard.edu/abs/2022ApJ...926..107M}
}

@ARTICLE{Ibata2020,
       author = {{Ibata}, Rodrigo and {Thomas}, Guillaume and {Famaey}, Benoit and {Malhan}, Khyati and {Martin}, Nicolas and {Monari}, Giacomo},
        title = "{Detection of Strong Epicyclic Density Spikes in the GD-1 Stellar Stream: An Absence of Evidence for the Influence of Dark Matter Subhalos?}",
      journal = {\apj},
         year = 2020,
        month = mar,
       volume = {891},
       number = {2},
          eid = {161},
        pages = {161},
          doi = {10.3847/1538-4357/ab7303},
archivePrefix = {arXiv},
       eprint = {2002.01488},
 primaryClass = {astro-ph.GA},
       adsurl = {https://ui.adsabs.harvard.edu/abs/2020ApJ...891..161I}
}

@ARTICLE{Ibata2024,
       author = {{Ibata}, Rodrigo and {Malhan}, Khyati and {Tenachi}, Wassim and {Ardern-Arentsen}, Anke and {Bellazzini}, Michele and {Bianchini}, Paolo and {Bonifacio}, Piercarlo and {Caffau}, Elisabetta and {Diakogiannis}, Foivos and {Errani}, Raphael and {Famaey}, Benoit and {Ferrone}, Salvatore and {Martin}, Nicolas F. and {di Matteo}, Paola and {Monari}, Giacomo and {Renaud}, Florent and {Starkenburg}, Else and {Thomas}, Guillaume and {Viswanathan}, Akshara and {Yuan}, Zhen},
        title = "{Charting the Galactic Acceleration Field. II. A Global Mass Model of the Milky Way from the STREAMFINDER Atlas of Stellar Streams Detected in Gaia DR3}",
      journal = {\apj},
         year = 2024,
        month = jun,
       volume = {967},
       number = {2},
          eid = {89},
        pages = {89},
          doi = {10.3847/1538-4357/ad382d},
archivePrefix = {arXiv},
       eprint = {2311.17202},
 primaryClass = {astro-ph.GA},
       adsurl = {https://ui.adsabs.harvard.edu/abs/2024ApJ...967...89I}
}

@ARTICLE{Belokurov2018,
       author = {{Belokurov}, V. and {Erkal}, D. and {Evans}, N.~W. and {Koposov}, S.~E. and {Deason}, A.~J.},
        title = "{Co-formation of the disc and the stellar halo}",
      journal = {\mnras},
         year = 2018,
        month = jul,
       volume = {478},
       number = {1},
        pages = {611-619},
          doi = {10.1093/mnras/sty982},
archivePrefix = {arXiv},
       eprint = {1802.03414},
 primaryClass = {astro-ph.GA},
       adsurl = {https://ui.adsabs.harvard.edu/abs/2018MNRAS.478..611B}
}

@ARTICLE{Ibata1994,
       author = {{Ibata}, R.~A. and {Gilmore}, G. and {Irwin}, M.~J.},
        title = "{A dwarf satellite galaxy in Sagittarius}",
      journal = {\nat},
         year = 1994,
        month = jul,
       volume = {370},
       number = {6486},
        pages = {194-196},
          doi = {10.1038/370194a0},
       adsurl = {https://ui.adsabs.harvard.edu/abs/1994Natur.370..194I}
}

@ARTICLE{Helmi2018,
       author = {{Helmi}, Amina and {Babusiaux}, Carine and {Koppelman}, Helmer H. and {Massari}, Davide and {Veljanoski}, Jovan and {Brown}, Anthony G.~A.},
        title = "{The merger that led to the formation of the Milky Way's inner stellar halo and thick disk}",
      journal = {\nat},
         year = 2018,
        month = oct,
       volume = {563},
       number = {7729},
        pages = {85-88},
          doi = {10.1038/s41586-018-0625-x},
archivePrefix = {arXiv},
       eprint = {1806.06038},
 primaryClass = {astro-ph.GA},
       adsurl = {https://ui.adsabs.harvard.edu/abs/2018Natur.563...85H}
}

@ARTICLE{Naidu2020,
       author = {{Naidu}, Rohan P. and {Conroy}, Charlie and {Bonaca}, Ana and {Johnson}, Benjamin D. and {Ting}, Yuan-Sen and {Caldwell}, Nelson and {Zaritsky}, Dennis and {Cargile}, Phillip A.},
        title = "{Evidence from the H3 Survey That the Stellar Halo Is Entirely Comprised of Substructure}",
      journal = {\apj},
         year = 2020,
        month = sep,
       volume = {901},
       number = {1},
          eid = {48},
        pages = {48},
          doi = {10.3847/1538-4357/abaef4},
archivePrefix = {arXiv},
       eprint = {2006.08625},
 primaryClass = {astro-ph.GA},
       adsurl = {https://ui.adsabs.harvard.edu/abs/2020ApJ...901...48N}
}

@ARTICLE{Pagnini2023,
       author = {{Pagnini}, G. and {Di Matteo}, P. and {Khoperskov}, S. and {Mastrobuono-Battisti}, A. and {Haywood}, M. and {Renaud}, F. and {Combes}, F.},
        title = "{The distribution of globular clusters in kinematic spaces does not trace the accretion history of the host galaxy}",
      journal = {\aap},
         year = 2023,
        month = may,
       volume = {673},
          eid = {A86},
        pages = {A86},
          doi = {10.1051/0004-6361/202245128},
archivePrefix = {arXiv},
       eprint = {2210.04245},
 primaryClass = {astro-ph.GA},
       adsurl = {https://ui.adsabs.harvard.edu/abs/2023A&A...673A..86P}
}

@ARTICLE{Johnston2002,
       author = {{Johnston}, Kathryn V. and {Spergel}, David N. and {Haydn}, Christian},
        title = "{How Lumpy Is the Milky Way's Dark Matter Halo?}",
      journal = {\apj},
         year = 2002,
        month = may,
       volume = {570},
       number = {2},
        pages = {656-664},
          doi = {10.1086/339791},
archivePrefix = {arXiv},
       eprint = {astro-ph/0111196},
 primaryClass = {astro-ph},
       adsurl = {https://ui.adsabs.harvard.edu/abs/2002ApJ...570..656J}
}

@ARTICLE{Ibata2002,
       author = {{Ibata}, R.~A. and {Lewis}, G.~F. and {Irwin}, M.~J. and {Quinn}, T.},
        title = "{Uncovering cold dark matter halo substructure with tidal streams}",
      journal = {\mnras},
         year = 2002,
        month = jun,
       volume = {332},
       number = {4},
        pages = {915-920},
          doi = {10.1046/j.1365-8711.2002.05358.x},
archivePrefix = {arXiv},
       eprint = {astro-ph/0110690},
 primaryClass = {astro-ph},
       adsurl = {https://ui.adsabs.harvard.edu/abs/2002MNRAS.332..915I}
}

@ARTICLE{Erkal2016,
       author = {{Erkal}, Denis and {Belokurov}, Vasily and {Bovy}, Jo and {Sanders}, Jason L.},
        title = "{The number and size of subhalo-induced gaps in stellar streams}",
      journal = {\mnras},
         year = 2016,
        month = nov,
       volume = {463},
       number = {1},
        pages = {102-119},
          doi = {10.1093/mnras/stw1957},
archivePrefix = {arXiv},
       eprint = {1606.04946},
 primaryClass = {astro-ph.GA},
       adsurl = {https://ui.adsabs.harvard.edu/abs/2016MNRAS.463..102E}
}

@ARTICLE{Erkal2017,
       author = {{Erkal}, Denis and {Koposov}, Sergey E. and {Belokurov}, Vasily},
        title = "{A sharper view of Pal 5's tails: discovery of stream perturbations with a novel non-parametric technique}",
      journal = {\mnras},
         year = 2017,
        month = sep,
       volume = {470},
       number = {1},
        pages = {60-84},
          doi = {10.1093/mnras/stx1208},
archivePrefix = {arXiv},
       eprint = {1609.01282},
 primaryClass = {astro-ph.GA},
       adsurl = {https://ui.adsabs.harvard.edu/abs/2017MNRAS.470...60E}
}

@ARTICLE{Erkal2019,
       author = {{Erkal}, D. and {Belokurov}, V. and {Laporte}, C.~F.~P. and {Koposov}, S.~E. and {Li}, T.~S. and {Grillmair}, C.~J. and {Kallivayalil}, N. and {Price-Whelan}, A.~M. and {Evans}, N.~W. and {Hawkins}, K. and {Hendel}, D. and {Mateu}, C. and {Navarro}, J.~F. and {del Pino}, A. and {Slater}, C.~T. and {Sohn}, S.~T. and {Orphan Aspen Treasury Collaboration}},
        title = "{The total mass of the Large Magellanic Cloud from its perturbation on the Orphan stream}",
      journal = {\mnras},
         year = 2019,
        month = aug,
       volume = {487},
       number = {2},
        pages = {2685-2700},
          doi = {10.1093/mnras/stz1371},
archivePrefix = {arXiv},
       eprint = {1812.08192},
 primaryClass = {astro-ph.GA},
       adsurl = {https://ui.adsabs.harvard.edu/abs/2019MNRAS.487.2685E}
}

@ARTICLE{White1978,
       author = {{White}, S.~D.~M. and {Rees}, M.~J.},
        title = "{Core condensation in heavy halos: a two-stage theory for galaxy formation and clustering.}",
      journal = {\mnras},
         year = 1978,
        month = may,
       volume = {183},
        pages = {341-358},
          doi = {10.1093/mnras/183.3.341},
       adsurl = {https://ui.adsabs.harvard.edu/abs/1978MNRAS.183..341W}
}

@ARTICLE{Ferrone2023,
       author = {{Ferrone}, Salvatore and {Di Matteo}, Paola and {Mastrobuono-Battisti}, Alessandra and {Haywood}, Misha and {Snaith}, Owain N. and {Montuori}, Marco and {Khoperskov}, Sergey and {Valls-Gabaud}, David},
        title = "{The e-TidalGCs project. Modeling the extra-tidal features generated by Galactic globular clusters}",
      journal = {\aap},
         year = 2023,
        month = may,
       volume = {673},
          eid = {A44},
        pages = {A44},
          doi = {10.1051/0004-6361/202244141},
archivePrefix = {arXiv},
       eprint = {2301.05166},
 primaryClass = {astro-ph.GA},
       adsurl = {https://ui.adsabs.harvard.edu/abs/2023A&A...673A..44F}
}

@ARTICLE{Ferrone2025,
       author = {{Ferrone}, Salvatore and {Montuori}, Marco and {Di Matteo}, Paola and {Mastrobuono-Battisti}, Alessandra and {Ibata}, Rodrigo and {Bianchini}, Paolo and {Khoperskov}, Sergey and {Leclerc}, Nicolas and {Hottier}, Clement and {Stein}, Eliot and {Valls-Gabaud}, David and {Owain Snaith}, N. and {Haywood}, Misha},
        title = "{Gaps in stellar streams as a result of globular cluster flybys: The case of Palomar 5}",
      journal = {\aap},
         year = 2025,
        month = jul,
       volume = {699},
          eid = {A289},
        pages = {A289},
          doi = {10.1051/0004-6361/202553923},
archivePrefix = {arXiv},
       eprint = {2502.03941},
 primaryClass = {astro-ph.GA},
       adsurl = {https://ui.adsabs.harvard.edu/abs/2025A&A...699A.289F}
}

@INPROCEEDINGS{NEMO,
       author = {{Teuben}, P.},
        title = "{The Stellar Dynamics Toolbox NEMO}",
    booktitle = {Astronomical Data Analysis Software and Systems IV},
         year = 1995,
       editor = {{Shaw}, R.~A. and {Payne}, H.~E. and {Hayes}, J.~J.~E.},
       series = {Astronomical Society of the Pacific Conference Series},
       volume = {77},
        month = jan,
        pages = {398},
       adsurl = {https://ui.adsabs.harvard.edu/abs/1995ASPC...77..398T}
}

@ARTICLE{Dehnen2002,
       author = {{Dehnen}, Walter},
        title = "{A Hierarchical <E10>O</E10>(N) Force Calculation Algorithm}",
      journal = {Journal of Computational Physics},
         year = 2002,
        month = jun,
       volume = {179},
       number = {1},
        pages = {27-42},
          doi = {10.1006/jcph.2002.7026},
archivePrefix = {arXiv},
       eprint = {astro-ph/0202512},
 primaryClass = {astro-ph},
       adsurl = {https://ui.adsabs.harvard.edu/abs/2002JCoPh.179...27D}
}

@ARTICLE{Yoon2011,
       author = {{Yoon}, Joo Heon and {Johnston}, Kathryn V. and {Hogg}, David W.},
        title = "{Clumpy Streams from Clumpy Halos: Detecting Missing Satellites with Cold Stellar Structures}",
      journal = {\apj},
         year = 2011,
        month = apr,
       volume = {731},
       number = {1},
          eid = {58},
        pages = {58},
          doi = {10.1088/0004-637X/731/1/58},
archivePrefix = {arXiv},
       eprint = {1012.2884},
 primaryClass = {astro-ph.GA},
       adsurl = {https://ui.adsabs.harvard.edu/abs/2011ApJ...731...58Y}
}

@ARTICLE{Carlberg2011,
       author = {{Carlberg}, R.~G. and {Richer}, Harvey B. and {McConnachie}, Alan W. and {Irwin}, Mike and {Ibata}, Rodrigo A. and {Dotter}, Aaron L. and {Chapman}, Scott and {Fardal}, Mark and {Ferguson}, A.~M.~N. and {Lewis}, G.~F. and {Navarro}, Julio F. and {Puzia}, Thomas H. and {Valls-Gabaud}, David},
        title = "{Density Variations in the NW Star Stream of M31}",
      journal = {\apj},
         year = 2011,
        month = apr,
       volume = {731},
       number = {2},
          eid = {124},
        pages = {124},
          doi = {10.1088/0004-637X/731/2/124},
archivePrefix = {arXiv},
       eprint = {1102.3501},
 primaryClass = {astro-ph.CO},
       adsurl = {https://ui.adsabs.harvard.edu/abs/2011ApJ...731..124C}
}

@ARTICLE{Sanders2016,
       author = {{Sanders}, Jason L. and {Bovy}, Jo and {Erkal}, Denis},
        title = "{Dynamics of stream-subhalo interactions}",
      journal = {\mnras},
         year = 2016,
        month = apr,
       volume = {457},
       number = {4},
        pages = {3817-3835},
          doi = {10.1093/mnras/stw232},
archivePrefix = {arXiv},
       eprint = {1510.03426},
 primaryClass = {astro-ph.GA},
       adsurl = {https://ui.adsabs.harvard.edu/abs/2016MNRAS.457.3817S}
}

@ARTICLE{Amorisco2016,
       author = {{Amorisco}, Nicola C. and {G{\'o}mez}, Facundo A. and {Vegetti}, Simona and {White}, Simon D.~M.},
        title = "{Gaps in globular cluster streams: giant molecular clouds can cause them too}",
      journal = {\mnras},
         year = 2016,
        month = nov,
       volume = {463},
       number = {1},
        pages = {L17-L21},
          doi = {10.1093/mnrasl/slw148},
archivePrefix = {arXiv},
       eprint = {1606.02715},
 primaryClass = {astro-ph.GA},
       adsurl = {https://ui.adsabs.harvard.edu/abs/2016MNRAS.463L..17A}
}

@ARTICLE{Pearson2017,
       author = {{Pearson}, Sarah and {Price-Whelan}, Adrian M. and {Johnston}, Kathryn V.},
        title = "{Gaps and length asymmetry in the stellar stream Palomar 5 as effects of Galactic bar rotation}",
      journal = {Nature Astronomy},
         year = 2017,
        month = aug,
       volume = {1},
        pages = {633-639},
          doi = {10.1038/s41550-017-0220-3},
archivePrefix = {arXiv},
       eprint = {1703.04627},
 primaryClass = {astro-ph.GA},
       adsurl = {https://ui.adsabs.harvard.edu/abs/2017NatAs...1..633P}
}

@ARTICLE{Banik2019,
       author = {{Banik}, Nilanjan and {Bovy}, Jo},
        title = "{Effects of baryonic and dark matter substructure on the Pal 5 stream}",
      journal = {\mnras},
         year = 2019,
        month = apr,
       volume = {484},
       number = {2},
        pages = {2009-2020},
          doi = {10.1093/mnras/stz142},
archivePrefix = {arXiv},
       eprint = {1809.09640},
 primaryClass = {astro-ph.GA},
       adsurl = {https://ui.adsabs.harvard.edu/abs/2019MNRAS.484.2009B}
}

@ARTICLE{Thomas2018,
       author = {{Thomas}, G.~F. and {Famaey}, B. and {Ibata}, R. and {Renaud}, F. and {Martin}, N.~F. and {Kroupa}, P.},
        title = "{Stellar streams as gravitational experiments. II. Asymmetric tails of globular cluster streams}",
      journal = {\aap},
         year = 2018,
        month = jan,
       volume = {609},
          eid = {A44},
        pages = {A44},
          doi = {10.1051/0004-6361/201731609},
archivePrefix = {arXiv},
       eprint = {1709.01934},
 primaryClass = {astro-ph.GA},
       adsurl = {https://ui.adsabs.harvard.edu/abs/2018A&A...609A..44T}
}

@ARTICLE{Thomas2023,
       author = {{Thomas}, Guillaume F. and {Famaey}, Benoit and {Monari}, Giacomo and {Laporte}, Chervin F.~P. and {Ibata}, Rodrigo and {de Laverny}, Patrick and {Hill}, Vanessa and {Boily}, Christian},
        title = "{Impact of the Galactic bar on tidal streams within the Galactic disc. The case of the tidal stream of the Hyades}",
      journal = {\aap},
         year = 2023,
        month = oct,
       volume = {678},
          eid = {A180},
        pages = {A180},
          doi = {10.1051/0004-6361/202346650},
archivePrefix = {arXiv},
       eprint = {2309.05733},
 primaryClass = {astro-ph.GA},
       adsurl = {https://ui.adsabs.harvard.edu/abs/2023A&A...678A.180T}
}

@ARTICLE{1DREAM,
       author = {{Canducci}, M. and {Awad}, P. and {Taghribi}, A. and {Mohammadi}, M. and {Mastropietro}, M. and {De Rijcke}, S. and {Peletier}, R. and {Smith}, R. and {Bunte}, K. and {Ti{\v{n}}o}, P.},
        title = "{1-DREAM: 1D Recovery, Extraction and Analysis of Manifolds in noisy environments}",
      journal = {Astronomy and Computing},
         year = 2022,
        month = oct,
       volume = {41},
          eid = {100658},
        pages = {100658},
          doi = {10.1016/j.ascom.2022.100658},
       adsurl = {https://ui.adsabs.harvard.edu/abs/2022A&C....4100658C}
}

@ARTICLE{Kupper2008,
       author = {{K{\"u}pper}, Andreas H.~W. and {MacLeod}, Andrew and {Heggie}, Douglas C.},
        title = "{On the structure of tidal tails}",
      journal = {\mnras},
         year = 2008,
        month = jul,
       volume = {387},
       number = {3},
        pages = {1248-1252},
          doi = {10.1111/j.1365-2966.2008.13323.x},
archivePrefix = {arXiv},
       eprint = {0804.2476},
 primaryClass = {astro-ph},
       adsurl = {https://ui.adsabs.harvard.edu/abs/2008MNRAS.387.1248K}
}

@ARTICLE{Kupper2010,
       author = {{K{\"u}pper}, Andreas H.~W. and {Kroupa}, Pavel and {Baumgardt}, Holger and {Heggie}, Douglas C.},
        title = "{Tidal tails of star clusters}",
      journal = {\mnras},
         year = 2010,
        month = jan,
       volume = {401},
       number = {1},
        pages = {105-120},
          doi = {10.1111/j.1365-2966.2009.15690.x},
archivePrefix = {arXiv},
       eprint = {0909.2619},
 primaryClass = {astro-ph.SR},
       adsurl = {https://ui.adsabs.harvard.edu/abs/2010MNRAS.401..105K}
}

@ARTICLE{Just2009,
       author = {{Just}, A. and {Berczik}, P. and {Petrov}, M.~I. and {Ernst}, A.},
        title = "{Quantitative analysis of clumps in the tidal tails of star clusters}",
      journal = {\mnras},
         year = 2009,
        month = jan,
       volume = {392},
       number = {3},
        pages = {969-981},
          doi = {10.1111/j.1365-2966.2008.14099.x},
archivePrefix = {arXiv},
       eprint = {0808.3293},
 primaryClass = {astro-ph},
       adsurl = {https://ui.adsabs.harvard.edu/abs/2009MNRAS.392..969J}
}

@ARTICLE{Pflamm-Altenburg23,
       author = {{Pflamm-Altenburg}, J. and {Kroupa}, P. and {Thies}, I. and {Jerabkova}, T. and {Beccari}, G. and {Prusti}, T. and {Boffin}, H.~M.~J.},
        title = "{Degree of stochastic asymmetry in the tidal tails of star clusters}",
      journal = {\aap},
         year = 2023,
        month = mar,
       volume = {671},
          eid = {A88},
        pages = {A88},
          doi = {10.1051/0004-6361/202244243},
archivePrefix = {arXiv},
       eprint = {2301.02251},
 primaryClass = {astro-ph.GA},
       adsurl = {https://ui.adsabs.harvard.edu/abs/2023A&A...671A..88P}
}

@ARTICLE{Pflamm-Altenburg2025,
       author = {{Pflamm-Altenburg}, J.},
        title = "{Asymmetry in the tidal tails of open star clusters from direct N-body integrations in Milgrom-law dynamics}",
      journal = {\aap},
         year = 2025,
        month = jan,
       volume = {693},
          eid = {A127},
        pages = {A127},
          doi = {10.1051/0004-6361/202347796},
archivePrefix = {arXiv},
       eprint = {2411.13675},
 primaryClass = {astro-ph.GA},
       adsurl = {https://ui.adsabs.harvard.edu/abs/2025A&A...693A.127P}
}

@ARTICLE{Kroupa2022,
       author = {{Kroupa}, Pavel and {Jerabkova}, Tereza and {Thies}, Ingo and {Pflamm-Altenburg}, Jan and {Famaey}, Benoit and {Boffin}, Henri M.~J. and {Dabringhausen}, J{\"o}rg and {Beccari}, Giacomo and {Prusti}, Timo and {Boily}, Christian and {Haghi}, Hosein and {Wu}, Xufen and {Haas}, Jaroslav and {Zonoozi}, Akram Hasani and {Thomas}, Guillaume and {{\v{S}}ubr}, Ladislav and {Aarseth}, Sverre J.},
        title = "{Asymmetrical tidal tails of open star clusters: stars crossing their cluster's pr{\'a}h$^{{\textdagger}}$ challenge Newtonian gravitation}",
      journal = {\mnras},
         year = 2022,
        month = dec,
       volume = {517},
       number = {3},
        pages = {3613-3639},
          doi = {10.1093/mnras/stac2563},
archivePrefix = {arXiv},
       eprint = {2210.13472},
 primaryClass = {astro-ph.GA},
       adsurl = {https://ui.adsabs.harvard.edu/abs/2022MNRAS.517.3613K}
}

@ARTICLE{Kroupa2024,
       author = {{Kroupa}, Pavel and {Pflamm-Altenburg}, Jan and {Mazurenko}, Sergij and {Wu}, Wenjie and {Thies}, Ingo and {Jadhav}, Vikrant and {Jerabkova}, Tereza},
        title = "{Open Star Clusters and Their Asymmetrical Tidal Tails}",
      journal = {\apj},
         year = 2024,
        month = jul,
       volume = {970},
       number = {1},
          eid = {94},
        pages = {94},
          doi = {10.3847/1538-4357/ad4c66},
archivePrefix = {arXiv},
       eprint = {2405.09609},
 primaryClass = {astro-ph.GA},
       adsurl = {https://ui.adsabs.harvard.edu/abs/2024ApJ...970...94K}
}

@ARTICLE{Weerasooriya25,
       author = {{Weerasooriya}, Sachi and {Starkenburg}, Tjitske and {Cunningham}, Emily C. and {Johnston}, Kathryn V},
        title = "{Dancing Streams In Merging Halos: Stellar Streams in a MW--LMC-like merger}",
      journal = {arXiv e-prints},
         year = 2025,
        month = may,
          eid = {arXiv:2505.14792},
        pages = {arXiv:2505.14792},
          doi = {10.48550/arXiv.2505.14792},
archivePrefix = {arXiv},
       eprint = {2505.14792},
 primaryClass = {astro-ph.GA},
       adsurl = {https://ui.adsabs.harvard.edu/abs/2025arXiv250514792W}
}

@ARTICLE{Besla2007,
       author = {{Besla}, Gurtina and {Kallivayalil}, Nitya and {Hernquist}, Lars and {Robertson}, Brant and {Cox}, T.~J. and {van der Marel}, Roeland P. and {Alcock}, Charles},
        title = "{Are the Magellanic Clouds on Their First Passage about the Milky Way?}",
      journal = {\apj},
         year = 2007,
        month = oct,
       volume = {668},
       number = {2},
        pages = {949-967},
          doi = {10.1086/521385},
archivePrefix = {arXiv},
       eprint = {astro-ph/0703196},
 primaryClass = {astro-ph},
       adsurl = {https://ui.adsabs.harvard.edu/abs/2007ApJ...668..949B}
}

@ARTICLE{Penarrubia2016,
       author = {{Pe{\~n}arrubia}, Jorge and {G{\'o}mez}, Facundo A. and {Besla}, Gurtina and {Erkal}, Denis and {Ma}, Yin-Zhe},
        title = "{A timing constraint on the (total) mass of the Large Magellanic Cloud}",
      journal = {\mnras},
         year = 2016,
        month = feb,
       volume = {456},
       number = {1},
        pages = {L54-L58},
          doi = {10.1093/mnrasl/slv160},
archivePrefix = {arXiv},
       eprint = {1507.03594},
 primaryClass = {astro-ph.GA},
       adsurl = {https://ui.adsabs.harvard.edu/abs/2016MNRAS.456L..54P}
}

@ARTICLE{Vasiliev2021,
       author = {{Vasiliev}, Eugene and {Belokurov}, Vasily and {Erkal}, Denis},
        title = "{Tango for three: Sagittarius, LMC, and the Milky Way}",
      journal = {\mnras},
         year = 2021,
        month = feb,
       volume = {501},
       number = {2},
        pages = {2279-2304},
          doi = {10.1093/mnras/staa3673},
archivePrefix = {arXiv},
       eprint = {2009.10726},
 primaryClass = {astro-ph.GA},
       adsurl = {https://ui.adsabs.harvard.edu/abs/2021MNRAS.501.2279V}
}

@ARTICLE{Dillamore2022,
       author = {{Dillamore}, Adam M. and {Belokurov}, Vasily and {Evans}, N. Wyn and {Price-Whelan}, Adrian M.},
        title = "{The impact of a massive Sagittarius dSph on GD-1-like streams}",
      journal = {\mnras},
         year = 2022,
        month = oct,
       volume = {516},
       number = {2},
        pages = {1685-1703},
          doi = {10.1093/mnras/stac2311},
archivePrefix = {arXiv},
       eprint = {2205.13547},
 primaryClass = {astro-ph.GA},
       adsurl = {https://ui.adsabs.harvard.edu/abs/2022MNRAS.516.1685D}
}

@ARTICLE{Bonaca2019,
       author = {{Bonaca}, Ana and {Hogg}, David W. and {Price-Whelan}, Adrian M. and {Conroy}, Charlie},
        title = "{The Spur and the Gap in GD-1: Dynamical Evidence for a Dark Substructure in the Milky Way Halo}",
      journal = {\apj},
         year = 2019,
        month = jul,
       volume = {880},
       number = {1},
          eid = {38},
        pages = {38},
          doi = {10.3847/1538-4357/ab2873},
archivePrefix = {arXiv},
       eprint = {1811.03631},
 primaryClass = {astro-ph.GA},
       adsurl = {https://ui.adsabs.harvard.edu/abs/2019ApJ...880...38B}
}

@ARTICLE{Mastrobuono-Battisti2012,
       author = {{Mastrobuono-Battisti}, A. and {Di Matteo}, P. and {Montuori}, M. and {Haywood}, M.},
        title = "{Clumpy streams in a smooth dark halo: the case of Palomar 5}",
      journal = {\aap},
         year = 2012,
        month = oct,
       volume = {546},
          eid = {L7},
        pages = {L7},
          doi = {10.1051/0004-6361/201219563},
archivePrefix = {arXiv},
       eprint = {1209.0466},
 primaryClass = {astro-ph.GA},
       adsurl = {https://ui.adsabs.harvard.edu/abs/2012A&A...546L...7M}
}

@BOOK{Spitzer1987,
       author = {{Spitzer}, Lyman},
        title = "{Dynamical evolution of globular clusters}",
         year = 1987,
       adsurl = {https://ui.adsabs.harvard.edu/abs/1987degc.book.....S}
}

@ARTICLE{FukushigeHeggie2000,
       author = {{Fukushige}, T. and {Heggie}, D.~C.},
        title = "{The time-scale of escape from star clusters}",
      journal = {\mnras},
         year = 2000,
        month = nov,
       volume = {318},
       number = {3},
        pages = {753-761},
          doi = {10.1046/j.1365-8711.2000.03811.x},
archivePrefix = {arXiv},
       eprint = {astro-ph/9910468},
 primaryClass = {astro-ph},
       adsurl = {https://ui.adsabs.harvard.edu/abs/2000MNRAS.318..753F}
}

@ARTICLE{Joshi2001,
       author = {{Joshi}, Kriten J. and {Nave}, Cody P. and {Rasio}, Frederic A.},
        title = "{Monte Carlo Simulations of Globular Cluster Evolution. II. Mass Spectra, Stellar Evolution, and Lifetimes in the Galaxy}",
      journal = {\apj},
         year = 2001,
        month = apr,
       volume = {550},
       number = {2},
        pages = {691-702},
          doi = {10.1086/319771},
archivePrefix = {arXiv},
       eprint = {astro-ph/9912155},
 primaryClass = {astro-ph},
       adsurl = {https://ui.adsabs.harvard.edu/abs/2001ApJ...550..691J}
}

@INCOLLECTION{Davies2013,
       author = {{Davies}, Melvyn B.},
        title = "{Globular Cluster Dynamical Evolution}",
    booktitle = {Planets, Stars and Stellar Systems. Volume 5: Galactic Structure and Stellar Populations},
         year = 2013,
       editor = {{Oswalt}, Terry D. and {Gilmore}, Gerard},
       volume = {5},
        pages = {879},
          doi = {10.1007/978-94-007-5612-0_17},
       adsurl = {https://ui.adsabs.harvard.edu/abs/2013pss5.book..879D}
}

@ARTICLE{Gibbons2014,
       author = {{Gibbons}, S.~L.~J. and {Belokurov}, V. and {Evans}, N.~W.},
        title = "{`Skinny Milky Way please', says Sagittarius}",
      journal = {\mnras},
         year = 2014,
        month = dec,
       volume = {445},
       number = {4},
        pages = {3788-3802},
          doi = {10.1093/mnras/stu1986},
archivePrefix = {arXiv},
       eprint = {1406.2243},
 primaryClass = {astro-ph.GA},
       adsurl = {https://ui.adsabs.harvard.edu/abs/2014MNRAS.445.3788G}
}

@ARTICLE{Bayer2025,
       author = {{Bayer}, M. and {Starkenburg}, E. and {Thomas}, G.~F. and {Martin}, N. and {Helmi}, A. and {Bystr{\"o}m}, A. and {de Boer}, T. and {Fern{\'a}ndez Alvar}, E. and {Gwyn}, S. and {Ibata}, R. and {Jablonka}, P. and {Kordopatis}, G. and {Matsuno}, T. and {McConnachie}, A.~W. and {Medina}, G.~E. and {S{\'a}nchez-Janssen}, R. and {Sestito}, F.},
        title = "{A Pristine-UNIONS view on the Galaxy: Kinematics of the distant spur feature of the Sagittarius stream traced by Blue Horizontal Branch stars}",
      journal = {arXiv e-prints},
         year = 2025,
        month = feb,
          eid = {arXiv:2502.17319},
        pages = {arXiv:2502.17319},
          doi = {10.48550/arXiv.2502.17319},
archivePrefix = {arXiv},
       eprint = {2502.17319},
 primaryClass = {astro-ph.GA},
       adsurl = {https://ui.adsabs.harvard.edu/abs/2025arXiv250217319B}
}

@ARTICLE{Koposov2023,
       author = {{Koposov}, Sergey E. and {Erkal}, Denis and {Li}, Ting S. and {Da Costa}, Gary S. and {Cullinane}, Lara R. and {Ji}, Alexander P. and {Kuehn}, Kyler and {Lewis}, Geraint F. and {Pace}, Andrew B. and {Shipp}, Nora and {Zucker}, Daniel B. and {Bland-Hawthorn}, Joss and {Lilleengen}, Sophia and {Martell}, Sarah L. and {S5 Collaboration}},
        title = "{S $^{5}$: Probing the Milky Way and Magellanic Clouds potentials with the 6D map of the Orphan-Chenab stream}",
      journal = {\mnras},
         year = 2023,
        month = jun,
       volume = {521},
       number = {4},
        pages = {4936-4962},
          doi = {10.1093/mnras/stad551},
archivePrefix = {arXiv},
       eprint = {2211.04495},
 primaryClass = {astro-ph.GA},
       adsurl = {https://ui.adsabs.harvard.edu/abs/2023MNRAS.521.4936K}
}

@ARTICLE{Baumgardt2018,
       author = {{Baumgardt}, H. and {Hilker}, M.},
        title = "{A catalogue of masses, structural parameters, and velocity dispersion profiles of 112 Milky Way globular clusters}",
      journal = {\mnras},
         year = 2018,
        month = aug,
       volume = {478},
       number = {2},
        pages = {1520-1557},
          doi = {10.1093/mnras/sty1057},
archivePrefix = {arXiv},
       eprint = {1804.08359},
 primaryClass = {astro-ph.GA},
       adsurl = {https://ui.adsabs.harvard.edu/abs/2018MNRAS.478.1520B}
}

@ARTICLE{Baumgardt2021,
       author = {{Baumgardt}, H. and {Vasiliev}, E.},
        title = "{Accurate distances to Galactic globular clusters through a combination of Gaia EDR3, HST, and literature data}",
      journal = {\mnras},
         year = 2021,
        month = aug,
       volume = {505},
       number = {4},
        pages = {5957-5977},
          doi = {10.1093/mnras/stab1474},
archivePrefix = {arXiv},
       eprint = {2105.09526},
 primaryClass = {astro-ph.GA},
       adsurl = {https://ui.adsabs.harvard.edu/abs/2021MNRAS.505.5957B}
}

@ARTICLE{Vasiliev&Baumgardt2021,
       author = {{Vasiliev}, Eugene and {Baumgardt}, Holger},
        title = "{Gaia EDR3 view on galactic globular clusters}",
      journal = {\mnras},
         year = 2021,
        month = aug,
       volume = {505},
       number = {4},
        pages = {5978-6002},
          doi = {10.1093/mnras/stab1475},
archivePrefix = {arXiv},
       eprint = {2102.09568},
 primaryClass = {astro-ph.GA},
       adsurl = {https://ui.adsabs.harvard.edu/abs/2021MNRAS.505.5978V}
}

@ARTICLE{Nibauer2025,
       author = {{Nibauer}, Jacob and {Bonaca}, Ana and {Price-Whelan}, Adrian M. and {Spergel}, David N. and {Greene}, Jenny E.},
        title = "{Measurement of Dark Matter Substructure from the Kinematics of the GD-1 Stellar Stream}",
      journal = {arXiv e-prints},
         year = 2025,
        month = oct,
          eid = {arXiv:2510.02247},
        pages = {arXiv:2510.02247},
          doi = {10.48550/arXiv.2510.02247},
archivePrefix = {arXiv},
       eprint = {2510.02247},
 primaryClass = {astro-ph.GA},
       adsurl = {https://ui.adsabs.harvard.edu/abs/2025arXiv251002247N}
}

@ARTICLE{Brooks2025,
       author = {{Brooks}, Richard A.~N. and {Garavito-Camargo}, Nicol{\'a}s and {Johnston}, Kathryn V. and {Price-Whelan}, Adrian M. and {Sanders}, Jason L. and {Lilleengen}, Sophia},
        title = "{LMC Calls, Milky Way Halo Answers: Disentangling the Effects of the MW{\textendash}LMC Interaction on Stellar Stream Populations}",
      journal = {\apj},
         year = 2025,
        month = jan,
       volume = {978},
       number = {1},
          eid = {79},
        pages = {79},
          doi = {10.3847/1538-4357/ad93a7},
archivePrefix = {arXiv},
       eprint = {2410.02574},
 primaryClass = {astro-ph.GA},
       adsurl = {https://ui.adsabs.harvard.edu/abs/2025ApJ...978...79B}
}

\begin{appendix}
\section{Length asymmetry for the separation into three populations of stellar streams} \label{Annexe_3_populations}

We classify the 36 stellar streams into three distinct groups based on the timing of their interaction with the merging satellite, which we define as the epoch when the stream, whether its progenitor or one of its tidal arms, reaches its closest approach to the satellite :

\begin{itemize} 
    \item[$\bullet$]  Streams perturbed before the merger event, between $t=1.7$ and $2.0$ Gyr: 5 streams on large orbits that were close to the infalling satellite.
    \item[$\bullet$] Streams affected during the satellite's first passage close to the Milky Way (MW) disk, around $t\approx2.2$ Gyr: 25 streams. Most of these are on orbits close to the MW; however, three are on wider orbits but were at pericentre at the time of the merger.
    \item[$\bullet$] Streams influenced after the satellite's first passage close to the MW disk, between $t=2.5$ and $3.3$ Gyr: 6 streams on large orbits. These experienced a deepening of the potential post-merger but were not directly impacted by the satellite.
\end{itemize}

\begin{figure}  
  \resizebox{\hsize}{!}{\includegraphics{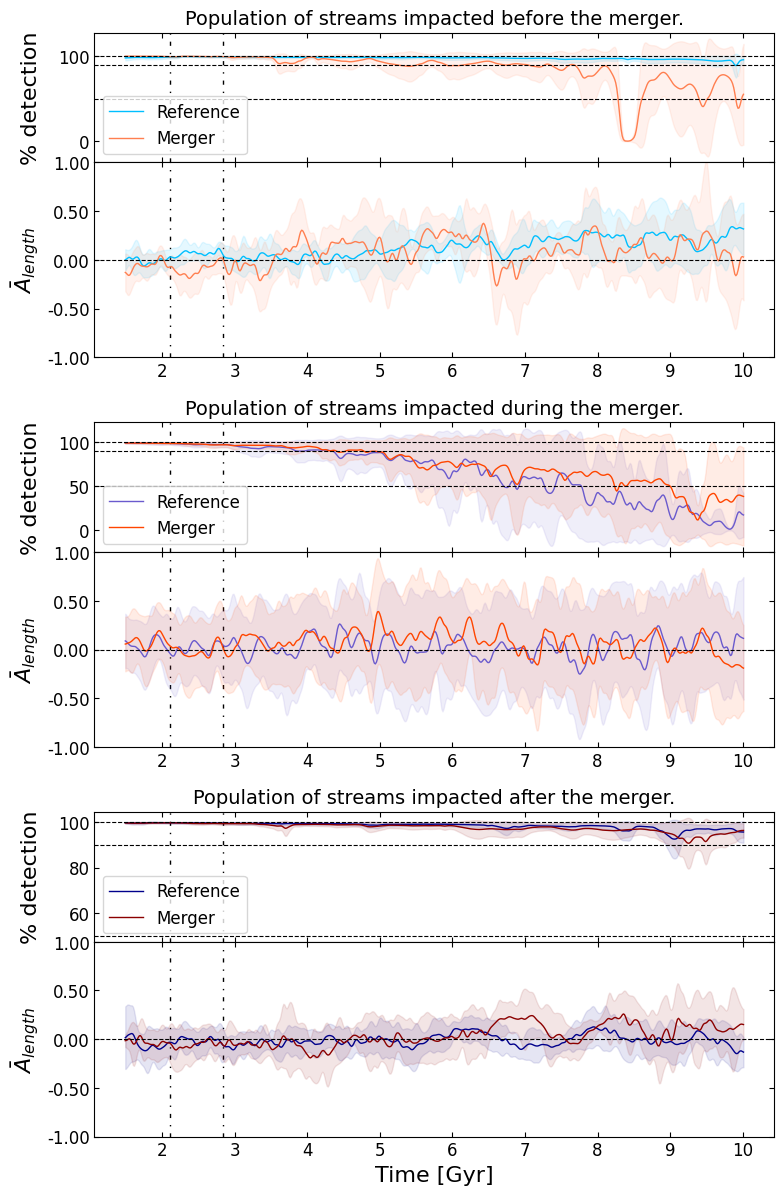}}
  \caption{Evolution of the median length asymmetry and detection fraction for three distinct populations of streams. \textit{Top row:} streams perturbed before the merger ($1.7$--$2.0$ Gyr). \textit{Middle row:} streams impacted during the first disk crossing ($\sim2.2$ Gyr). \textit{Bottom row:} streams influenced after the first disk crossing ($2.5$--$3.3$ Gyr).}
  \label{Length_asymmetry_3_pop}
\end{figure}

Fig.~\ref{Length_asymmetry_3_pop} shows, for each population, the median length asymmetry and the fraction of particles detected by our algorithm. The first-passage group (middle row) shows no clear asymmetry in this plot. This may originate from our difficulty in measuring the arm lengths: when on orbits close to the MW, strong tidal forces can shred the streams, and make many particles undetectable, lowering the detection fraction. Moreover, any asymmetry arising from the merging event in one stream would be washed out when averaging over the population, due to unsynchronized oscillations.

By contrast, streams in the pre- and post-first-passage groups (top and bottom rows) on wide orbits remain well detected by the algorithm, maintaining detection fractions above $90\%$. In the pre-first-passage sample (top row), the $1\sigma$ scatter (shaded envelope) around $t=2.0$ Gyr increases relative to the reference simulation, indicating that the satellite induces an asymmetry. Afterward, this scatter exhibits oscillations driven by the asymmetry caused by the eccentric orbit in each stream and the lack of phase coherence across the population. Despite the unsynchronized oscillations of the population, which occur in both runs, the scatter seems to remain larger in the merger case than in the reference case. In the post-first-passage sample (bottom row), the scatter increases slightly after 3.5 Gyr and persists for the rest of the simulation.

Most streams, especially those near the Milky Way, do not exhibit a clear merger-induced length asymmetry. Eccentric, unsynchronized phases obscure any signal, and rapid disruption prevents reliable length recovery by 1-DREAM. Consequently, the modest signal retained by a minority of wide-orbit streams is washed out when averaging over the full population (see Fig.~\ref{Length_asymmetry}).

\section{Asymmetries in number of particles and in density.} \label{Annexe_asymmetry_nb_particules+density}

 In Section \ref{need_new_asym} we note that asymmetry in density proved to not be suited for detecting merger signatures. Here we describe those tests in detail.

For each stream arm at time $t$, we compute the normalized particle‐count asymmetry
\begin{equation}
    q_{nb} = \frac{N_l-N_t}{N_l+N_t}
\end{equation}
where $N_l$ and $N_t$ are the numbers of particles in the leading and trailing arms, respectively \citep[corresponding to the $\epsilon$ quantity in ][]{Pflamm-Altenburg23, Kroupa2024, Pflamm-Altenburg2025}, and an analogous quantity $q$ restricted to the interval 2-15 kpc interval from the progenitor. We recall (Section \ref{1-dream}) that the stream is binned into 1 kpc segments starting at 2 kpc from the progenitor, ensuring no contamination from the progenitor and focusing solely on the tidal arms.

\begin{figure}  
  \resizebox{\hsize}{!}{\includegraphics{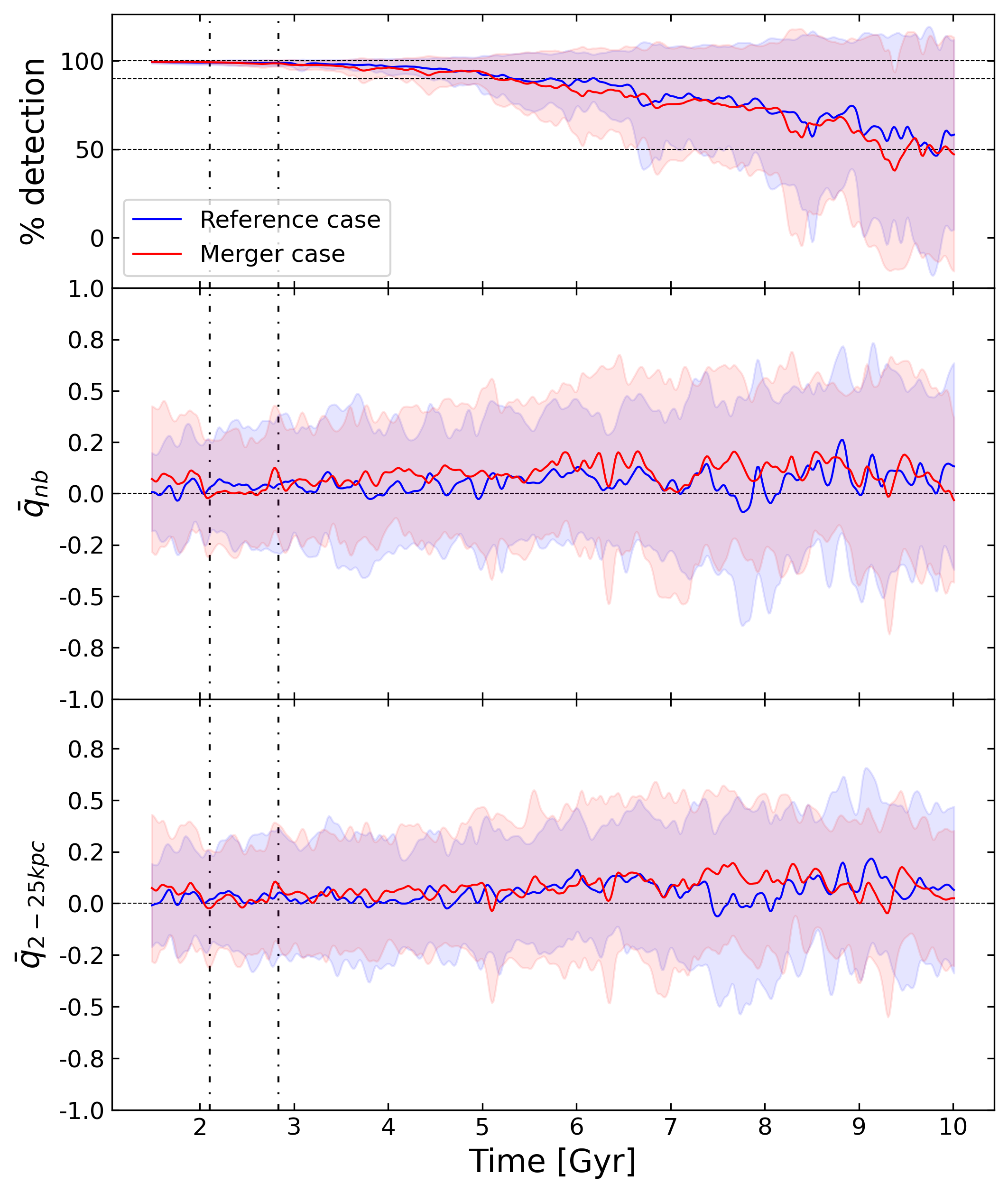}}
  \caption{\textit{Top row:} Fraction of stream particles recovered by the 1-DREAM algorithm as a function of time. \textit{Middle row:} Median asymmetry of the number of particles detected between the leading and trailing arm; shaded area indicate the $\pm1\sigma$ scatter. \textit{Bottom row:} Same but the asymmetry is computed in the range $2-15$ kpc.}
  \label{asymmetry_nb_particules+density}
\end{figure}

Fig.~\ref{asymmetry_nb_particules+density} shows the median over the population of stellar streams and $1\sigma$ envelope of $q_{nb}$ (middle panel) and $q_{2-15\text{kpc}}$ (bottom panel) for both the reference and merger runs, along with the fraction of stream particles recovered by the 1-DREAM algorithm (top panels). Although the merger run exhibits a modest increase in scatter for both metrics, this coincides with a decrease in detection fraction, indicating the effect may come from numerical artefacts in particle recovery rather than a true merger imprint.

These diagnostics are therefore not robust merger tracers in our setup and motivate the cumulative-density metric used in the main text.

\end{appendix}

\end{document}